%% file: NUHM_CNMSSM_v3.tex
\documentclass[11pt]{article}

\usepackage{calc}
\usepackage{rotating}
\usepackage[english]{babel}
\usepackage{graphicx}
\usepackage{subfig}
\usepackage{float}
\usepackage{amsmath}
\usepackage{amssymb}
\usepackage{amsthm}
\usepackage{latexsym}
\usepackage{dcolumn}
\usepackage{geometry}
\usepackage{color}
\usepackage{hyperref}
\usepackage{mciteplus}

\def\gtwid{\mathrel{\raise.3ex\hbox{$>$\kern-.75em\lower1ex\hbox{$\sim
$}}}}
\def\vio{\mathrel{\hbox{$E$\kern-.60em\hbox{$/
$}}}}
\input macros_bf4

\begin{document}

\title{Simultaneous enhancement in $\gamma\gamma,\;b\bar{b}$ and
  $\tau^+\tau^-$ rates in the NMSSM with
 nearly degenerate scalar and pseudoscalar Higgs bosons}

\author{\bf Shoaib Munir,~Leszek Roszkowski\footnote{On leave of absence
    from the University of Sheffield, UK.},~Sebastian Trojanowski\\
National Centre for Nuclear Research,
  Ho{\. z}a 69, 00-681 Warsaw, Poland \\
\bigskip \\
\url{Shoaib.Munir@fuw.edu.pl},\\ \url{L.Roszkowski@sheffield.ac.uk},\\
\url{Sebastian.Trojanowski@fuw.edu.pl}} 

\maketitle

 \begin{abstract} 
   We propose an experimental test of a scenario in the
   Next-to-Minimal Supersymmetric Standard Model in which both the
   lightest scalar and the lightest pseudoscalar Higgs bosons have masses around
   125\gev. The pseudoscalar can contribute significantly to the
   $\gamma\gamma$ rate at the LHC due to light Higgsino-like charginos
   in its effective one-loop coupling to two photons. Such charginos
   are obtained for small values of the $\mu_{\rm eff}$ parameter
   which also results in enhanced $b\bar{b}$ and $\tau^+\tau^-$ rates
   compared to those expected for a Standard Model Higgs boson.  This
   scenario should result in a clear discrepancy between the observed
   rates in these three decay channels and those in the $WW$ and $ZZ$
   channels, since the pseudoscalar does not couple to the $W$ and $Z$
   bosons.  However, in the dominant gluon fusion production mode the
   pseudoscalar will stay hidden behind the SM-like scalar Higgs boson
   and in order for it to be observable the associated $b\bar{b}h$
   production mode has to be considered, the cross section for which
   is tiny in the Standard
   Model but $\tanb$-enhanced in supersymmetry. We analyze the
   constrained NMSSM with non-universal Higgs sector parameters and
   identify regions of its parameter space where the lightest
   pseudoscalar with mass around 125\gev\ and strongly enhanced $\gamma\gamma$
   (up to 60\%), $b\bar{b}$ and $\tau^+\tau^-$ rates in the
   $b\bar{b}h$ mode can be obtained.
\end{abstract}

\section{\label{intro}Introduction}

Since its discovery at the LHC in
July 2012\cite{Chatrchyan:2012ufa,Aad:2012tfa}, the CMS and ATLAS 
collaborations have accumulated
more data and updated their results on the Higgs boson. In the
early results, a considerable
enhancement in the $\gamma\gamma$ and $ZZ$ rates 
compared to the Standard Model (SM) prediction was noted near
$\sim$125\gev\ at the ATLAS detector. According to the CMS data,
the signal strength, $\sigma/\sigma_{\rm SM}$, was consistent with the SM prediction in the
$ZZ$ channel but an enhancement in the $\gamma\gamma$ channel was
observed there as well. However, the figures from both experiments have changed in the
latest results released after the collection of $\sim$20/fb of
data\cite{CMS-H-twiki,ATLAS-H-twiki}. The signal strengths measured
by the CMS have now fallen down to SM-like values, $0.78\pm 0.27$
 and $0.91^{+0.30}_{-0.24}$ in the
 $\gamma\gamma$ and $ZZ$ decay channels, respectively, with the mean value of the
 boson mass being $125.6\pm0.64$\gev. 

The ATLAS collaboration, on the
 other hand, still reports sizable excesses, $\sigma/\sigma_{\rm SM} =1.65\pm0.35$ in the
 $\gamma\gamma$ channel with the mass measurement yielding
 $126.8\pm0.73$\gev, 
and $\sigma/\sigma_{\rm SM}=1.7\pm0.5$ in the $ZZ$ channel with mass at
 $124.3^{+0.55}_{-0.4}$\gev. Moreover, broad peaks consistent with a
125\gev\ boson have now also been observed in the $H\rightarrow WW
\rightarrow 2l2\nu$ channel in the two detectors. 
Importantly, the best-fit signal strength in this channel
is SM-like according to both CMS and ATLAS,
with a measured value of $0.76\pm0.21$ at the former and of $1.01\pm
0.31$ at the latter. In the $b\bar{b}$ decay channel, 
  although no significant excess has been observed above the SM
  background at either CMS or ATLAS, fitted
  signal strength values at $m_h=125$\gev\ have been obtained by the
  two collaborations for some individual Higgs boson production modes. The
  best-fit values provided by the CMS collaboration read $0.7\pm 1.4$
  for vector boson fusion (VBF), $1.0^{+0.5}_{-0.5}$ for Higgs-strahlung off a vector boson
($Wh/Zh$) and $0.74^{+1.34}_{-1.30}$ for associated
production off top quarks ($t\bar{t}h$). The ATLAS collaboration has
recently provided a fitted value of $0.2\pm 0.64$ for the
$Wh/Zh$ production mode only. In the $\tau^+\tau^-$ channel an excess of events over
  a broad $m_h$ range was reported by the CMS collaboration with a
  best-fit $\sigma/\sigma_{\rm SM} = 1.1\pm 0.4$ at 125\gev. At ATLAS,
  however, no excess has so far been observed in this channel also and
  a fitted value at $m_h=125$\gev, which is SM-like but with a very large
  error, can be noted in\cite{ATLAS-CONF-2013-034}. We should point
  out here that neither of the two collaborations have provided any
  measurements of the signal strength in 
  the $b\bar{b}$ and $\tau^+\tau^-$ decay channels for associated Higgs 
 boson production off bottom quarks ($b\bar{b}h$).

Since the first announcements of the discovery of the boson, there
have been many attempts to interpret the observed data in light of
various supersymmetric (SUSY) extensions of the
SM\cite{Chung:2012vg,*Baer:2012mv,*Heng:2012at,*Carena:2012mw,*Bechtle:2012jw,*Berenstein:2012fc,*Cheung:2012pq,*Cao:2013ur,Fowlie:2012im,Kowalska:2013hha,King:2012is,*Cao:2012fz,*Vasquez:2012hn,*Jeong:2012ma,*Rathsman:2012dp,*Das:2012ys,*Cheng:2013,*Cao:2012yn,Gunion:2012gc,Bae:2012am,*Cheng:2012pe,*Kang:2012bv,
  *Perelstein:2012qg,*SchmidtHoberg:2012yy,*Belanger:2012sd,Gunion:2012he,Agashe:2012zq,*Chalons:2012qe,*Gogoladze:2012jp,*Belanger:2012tt,
  *Dreiner:2012ec,*King:2012tr,*Choi:2012he,*Gherghetta:2012gb,*LopezFogliani:2012yq,*Cao:2013si,*Das:2013ta,*Cerdeno:2013cz,*Christensen:2013dra,*Wang:2013jya,*Barbieri:2013hxa,*Moretti:2013lya,Kowalska:2012gs,Athron:2012sq,*Basso:2012tr,*SchmidtHoberg:2012ip,*Benakli:2012cy}. In
the context of the Minimal Supersymmetric Standard Model (MSSM) the
observed signal can be interpreted as being due to the lightest Higgs
boson of the model, $h$. In the MSSM constrained at the grand
unification theory (GUT) scale, referred to as the
CMSSM\cite{hep-ph/9312272}, $h$ can attain a mass around the measured
central value only if the SUSY-breaking scale, \msusy, is close to or
larger than 1\tev, while also satisfying other important
phenomenological constraints. In the Next-to-Minimal Supersymmetric
Standard Model (NMSSM)\cite{Drees:1988fc,PhysRevD.39.844} (see,
e.g.,\cite{Ellwanger:2009dp,Maniatis:2009re} for recent reviews) it has been
shown that either of the two lightest CP-even Higgs bosons, $h_1$ and
$h_2$, can easily be SM-like with mass around
125\gev\cite{Ellwanger:2011aa,Benbrik:2012rm}. In fact in this model
it is possible to have $h_1$ and $h_2$ almost degenerate in mass
around 125\gev \cite{Gunion:2012gc}, so that the observed signal is
actually a superposition of two individual peaks due to each of these,
which cannot be independently resolved.

In the GUT-constrained version of the NMSSM
(CNMSSM)\cite{PhysRevD.39.844,Djouadi:2008yj,Djouadi:2008uj,Gunion:2012zd},
in analogy with the CMSSM,
it has been found that in order to obtain $h_1$ as heavy
as 125\gev, \msusy\ at or above 1\tev\ is needed even with relevant phenomenological
constraints imposed\cite{Kowalska:2012gs}. Alternatively, a SM-like
$h_2$ with mass $\sim$125\gev\ is easily
achievable\cite{Kowalska:2012gs}. Relaxing slightly the
universality conditions by disunifying the masses of the scalar
Higgs doublets $m_{H_u}$ and $m_{H_d}$ from the
scalar mass parameter $m_0$ and the soft Higgs trilinear
coupling parameters \alam\ and \akap\ from the unified soft Yukawa coupling \azero, 
makes it relatively easy to obtain SM-like 
$h_1$ or $h_2$ around 125\gev\cite{Ellwanger:2012ke}. Here we refer to such a model with
non-universal Higgs sector parameters as CNMSSM-NUHM. The scenario with mass degenerate
$h_1$ and $h_2$ satisfying also other phenomenological constraints has
also been pursued with much interest in the CNMSSM-NUHM\cite{Gunion:2012gc,Gunion:2012he}.

Even though the latest results from CMS seem to favor a SM-like Higgs boson, those from
ATLAS do so only partially and
it is still possible for the observed boson to be a non-standard one. The inconsistencies
between the various measurements and fluctuations in the data
leave ample room for speculation in this regard. 
Therefore, in this article, we propose an experimental test of a scenario, not investigated
hitherto, in which the lightest pseudoscalar, $a_1$,  of the NMSSM
is almost degenerate in mass with the lightest
$\sim$125\gev\ scalar, $h_1$. Such $a_1$ can have a sizeable  
one-loop effective coupling to $\gamma\gamma$ in the presence of a light
Higgsino-like chargino in the loop.
Thus with decreasing mass of such a chargino one should expect a rise in the signal rate
defined, for a general Higgs boson $h_i$, as
\begin{eqnarray}
\label{eq:Rxsct1}
R^Y_X(h_i) =  \frac{\sigma(Y \rightarrow
  h_i)}{\sigma(Y \rightarrow h_{\rm SM})}\times \frac{BR(h_i
  \rightarrow X)}{BR(h_{\rm SM} \rightarrow X)}\,,
\end{eqnarray}
where $h_{\rm SM}$ is a SM Higgs boson with
the same mass as $h_i$ and $X$ denotes any one of its allowed SM decay
channels. $Y$ stands for the various possible Higgs boson production modes at the LHC, which include gluon fusion
($ggh$), VBF, $Wh/Zh$\footnote{We note here that the VBF and $Wh/Zh$ production
  modes are irrelevant for a pseudoscalar Higgs boson.} and $t\bar{t}h/b\bar{b}h$. 
However, for $X=\gamma\gamma$, despite the non-negligible size of the second
term in the product on the right hand side of Eq.\,(\ref{eq:Rxsct1}),
no net enhancement in the $\gamma\gamma$ rate of $a_1$ with decreasing
chargino mass would be visible in the $ggh$
production mode. The reason is that 
the first term in the product always has a very small magnitude due to a
highly reduced effective coupling of $a_1$ to
two gluons compared to that of a SM Higgs boson, which is dominated by 
the top quark loop, thus nullifying the overall effect. 

The overall enhancement in $R^Y_X(a_1)$ due to a light chargino should instead be visible in the $b\bar{b}h$
production mode since, as we shall see, the conditions necessary to obtain a
light chargino also result in an enhanced coupling of $a_1$ to
$b\bar{b}/\tau^+\tau^-$. In fact, one should thus obtain a
simultaneous enhancement in
the signal rates of the three channels, $\gamma\gamma$, $b\bar{b}$ and
$\tau^+\tau^-$ (collectively referred to as $X$ henceforth). We point
out here that, while the enhancement in the $b\bar{b}$ and $\tau^+\tau^-$ channels only 
is in principle possible even with a light MSSM-like scalar Higgs boson for large
$\tan\beta$, the above `triple enhancement' should be a clear signature of our proposed scenario.
For this reason we shall investigate $b\bar{b}h$ Higgs production mode
here, emphasizing the importance of a measurement of the signal
rate in this mode, which
is very  subdominant for a SM Higgs boson and is therefore
generally considered to be of less interest. In contrast, in SUSY
it is enhanced by $\tan^2\beta$\cite{Dawson:2005vi,*Dawson:2011jy}
and can therefore be potentially very interesting.

In the $b\bar{b}h$ channel, the $a_1$ could be partially responsible
for a net enhancement in the signal rate, $R^{bb}_X$(obs), in the
$X$ decay channels measured at the LHC. However, being a pseudoscalar, 
it would not contribute to the $WW$ and $ZZ$ channels
(denoted collectively by $V$), so that, assuming $h_1$ to be exactly SM-like,
\begin{equation}
R^{bb}_X ({\rm obs}) \equiv \frac{\sigma^{bb}_X({\rm obs})}{\sigma^{bb}_X (h_{\rm
    SM})} = R^{bb}_X (h_1) +
R^{bb}_X (a_1)\simeq
1+ R^{bb}_X (a_1) \ \ \
\textrm{and}  \ \ \ R^{bb}_V
({\rm obs}) = R^{bb}_V (h_1) \simeq 1\,.
\end{equation}
Furthermore, a difference in the mass measurements 
in the $X$ and $V$ modes would also provide a hint for
mass degenerate ${h_1}$ and $a_1$. Such a degeneracy would imply that 
the signal observed in the $X$ channels should in fact be
interpreted as the `sum' of two individual peaks due to $h_1$ and $a_1$,
while the peaks in the $V$ modes correspond to $h_1$ alone. $h_1$ is
still SM-like in this scenario due to a significant singlet component
even though $\tan\beta$ can take fairly large values\cite{Badziak:2013bda}.
Since this scenario is compatible with a SM-like scalar Higgs boson, it is
also not in conflict with the recent CMS
measurements in the $ZZ$ mode which 
disfavor the pure pseudoscalar hypothesis\cite{Chatrchyan:2012jja,Djouadi:2013qya}. 

We identify regions of the CNMSSM-NUHM parameter space where both
$h_1$ and $a_1$ with masses around 125\gev\ can be obtained, expecting
that a discrepancy between $X$ and $V$ rates will be seen by CMS and
ATLAS collaborations in a focussed analysis of the $b\bar{b}h$
production mode.  We further confine ourselves only to the regions
where the above mentioned triple enhancement can be obtained, serving
as a clear signature of this scenario. We investigate the impact of
other important experimental constraints on these regions.
These include the limits from direct SUSY searches released by ATLAS with
$\sim$20/fb of data as well as from the dark matter (DM) relic density measurements. 
We also require the corresponding parameter space to satisfy the
recently announced positive \brbsmumu\ measurement by the LHCb
collaboration.

The article is organized as follows. In \refsec{Pseudo} we discuss the
possibility of observing an enhancement in the $\gamma\gamma$ and $b\bar{b}/\tau^+\tau^-$ rates at
the LHC due to a $\sim$125\gev\ pseudoscalar Higgs boson. In  
\refsec{Model} we define the model's parameter space. In \refsec{Results} we
describe the experimental constraints
applied in our scans, present our numerical results and discuss their salient
features. We summarize our findings in \refsec{Summary}.


\section{\label{Pseudo}Enhancement in the observed $\gamma\gamma$ rate due to a light pseudoscalar}

In this section we present some analytical details of the mentioned
NMSSM scenario in which the correlation between the $\gamma\gamma$ and $WW/ZZ$
rates can be altered. 
One way to achieve this is with mass degenerate lightest doublet-like
scalar Higgs, $h_1$, and lightest singlet-like pseudoscalar, $a_1$.

\subsection{The pseudoscalar mass}
\label{sec:a1mass}

We first discuss the conditions that are necessary to obtain a $\sim$125 singlet-like
$a_1$ which couples to two photons through loops of fermions and
charginos only. Starting from the $2\times 2$
pseudoscalar mass matrix (after rotating away the Goldstone
mode)\cite{Ellwanger:2009dp}, one can obtain the approximate expression,
\begin{equation}
\label{eq:ma1}
m_{a_1}^2 \simeq -3\kappa sA^{\rm SUSY}_{\kappa} - \frac{M_{P,12}^4}{M_{P,11}^2}\,.
\end{equation}
In the above equation $M_{P,12}^2 \simeq \lambda(A^{\rm
  SUSY}_{\lambda} - 2\kappa s)v$ is the off-diagonal entry of the
pseudoscalar mass matrix, where $v \equiv
\sqrt{v_u^2 + v_d^2} \simeq 174$\gev, with $v_u$ and $v_d$ being the
vacuum expectation values (vevs) of the $u$-type and $d$-type Higgs doublets,
respectively, and $A^{\rm SUSY}_{\lambda/\kappa}$ denoting
$A_{\lambda/\kappa}$ at \msusy. $M_{P,11}^2 \simeq \mueff B_{\rm eff}\tan\beta$, with
$\mueff \equiv \lambda s$ ($s$ being the vev of the singlet field $S$), $B_{\rm eff} \equiv
A^{\rm SUSY}_{\lambda} + \kappa s$ and $\tan\beta \equiv v_u/v_d$, is
the diagonal term corresponding to the mass-squared of the doublet-like heavy
pseudoscalar, $a_2$. The leading term in Eq.\,(\ref{eq:ma1}) implies
that, for positive \kap, which we will assume here, the condition
of the positivity of $m_{a_1}^2$ depends predominantly on the relative
signs of \mueff\ and \akap\ at \msusy. This condition thus has some
important repercussions when \akap\ and \alam\ are taken as input
parameters at the GUT scale. 
Assuming the leading term to be positive so that the correct $m_{a_1}$ is achieved by adjusting
the free parameters in it, the negative contribution from the second
term should be kept close to zero. This would require
$M_{P,11}^2\gtrsim M_{P,12}^4$. We explain how this can be achieved for negative and
positive \mueff\ in the following.  

For $\mueff < 0$ (and therefore negative $s$, assuming positive \lam),
the first term in Eq.\,(\ref{eq:ma1}) is positive if $\akap>0$ at \msusy.
In the second term $M_{P,12}^4$ is positive definite
and $M_{P,11}^2$ must be positive for
a non-tachyonic $a_2$, which requires $B_{\rm eff} < 0$. 
For given $\tan\beta$
and $\mueff$, $M_{P,11}^2$ is driven by the magnitude of $B_{\rm
  eff}$, in order to enhance which \alam\ should take smaller values
(note that \alam\ is bounded from above
by $\kappa|s|$).  
However, \alam\ at \msusy\ runs upwards from its GUT value with
falling negative \azero\ owing to the
contribution from the relevant term in its
renormalization group equation (RGE)\cite{Ellwanger:2009dp}. Hence increasing negative
\azero\ diminishes the difference
between the two terms in $B_{\rm eff}$, reducing its
size and in turn driving $M_{P,11}^2$ closer to zero. At the same time
$M_{P,12}$, which is a sum of $2\kappa |s|$ and \alam,
grows as \alam\ increases, as opposed to $M_{P,11}^2$. 
Consequently, the ratio
$\frac{M_{P,12}^4}{M_{P,11}^2}$ in Eq.\,(\ref{eq:ma1}) grows with decreasing \azero\ and,
for large negative values of the latter, can result in negative
$m_{a_1}^2$. Note also that the running of \akap\ in turn depends dominantly on \alam. \akap\ runs upwards with
\alam\ as long as the latter is negative. When \alam\ turns positive
\akap\ runs in the opposite direction, owing to its
RGE. 
Thus \akap\ in the leading term in
Eq.\,(\ref{eq:ma1}) will have somewhat constrained GUT scale values  
that can yield correct $m_{a_1}$.
On the other hand, for $\mueff > 0$, the two terms in $B_{\rm eff}$ are both positive and
 the cancellation described above does not occur. 

 In summary, the net effect of the interplay between various Higgs
 sector parameters is that for negative \mueff\ the values of \azero\
 at the GUT scale are bounded from below by the condition of the
 physicality of $a_1$. This constraint on \azero\ causes a slight
 tension between $m_{h_1}$ and $m_{a_1}$, since it is well known that
 in order to obtain $h_1$ which is SM-like with mass $\sim$125\gev\
 large negative values of \azero\ are required for
 $\msusy\sim1\tev$. For positive \mueff\ there is no such tension
 because \azero\ is relatively free to take values that give large
 negative $A_t$ at \msusy, as long as the correct $a_1$ mass can be
 achieved by adjusting other free parameters.

\subsection{$\gamma\gamma$ decay of the pseudoscalar}
\label{sec:a1gamgam}
Besides a singlet-like $a_1$ with mass similar to that of the
experimentally observed boson, this scenario
also requires a low mass, $m_{\chi^{\pm}_1}$, of the lightest chargino.
The effective coupling of a pseudoscalar $a_i,$ with $i=1,2$, to two
photons (see, e.g.,\cite{Spira:1997dg,Djouadi:2005gj}), is dominated
by a light chargino in the loops and can be
approximated by
\begin{equation}
\label{eq:eff-cpl}
C^{\rm eff}_{a_i} (\gamma\gamma) \simeq
\frac{g_{a_1\chi^{\pm}_1\chi^{\pm}_1}}{\sqrt{\sqrt{2}G_F}\,m_{\chi^{\pm}_1}}\,A_{1/2}^{a_i}(\tau_i)\,,
\end{equation}
where $\tau_i = \frac{m_{a_i}^2}{4m^2_{\chi_1^\pm}}$. For $\tau_i \le
1$, which is applicable here, with $m_{a_i} \simeq 126$\gev\ and the light chargino obeying the
lower limit, $m_{\chi^{\pm}_1} > 94$\gev\cite{Beringer:1900zz}, the form-factor 
$A_{1/2}^{a_i} (\tau_i)
=\frac{1}{\tau_i}\arcsin^2{\sqrt{\tau_i}}$\cite{Djouadi:2005gi} 
in the above equation lies in the range
\begin{equation}
1 < A_{1/2}^{a_i} (\tau_i) \lesssim 1.2.
\end{equation}
 The coupling of $a_i$ to charginos in Eq.\,(\ref{eq:eff-cpl}) can be
 written, following the notation of\cite{Ellwanger:2009dp}, as 
\begin{equation}
\label{eq:coupling}
g_{a_i\chi^{\pm}_1\chi^{\pm}_1} = i\Big[\frac{\lambda}{\sqrt{2}}\,P_{i3}\,\sin\theta_U\,\sin\theta_V - \frac{g_2}{\sqrt{2}}(P_{i2}\,\cos\theta_U\,\sin\theta_V + P_{i1}\,\sin\theta_U\,\cos\theta_V)\Big]\,,
\end{equation}
where $\theta_U$, $\theta_V$ are the mixing angles for rotating the
chargino interaction states to mass eigenstates, and $P_{ij}$ are the
entries of the mixing matrix that diagonalizes the pseudoscalar mass
matrix. When the pseudoscalar weak eigenstates $A_i^{\rm weak}$ are
expressed in the basis
($H_{dI},\,H_{uI},\,S_I$)\cite{Ellwanger:2009dp}, $P_{i1}$ corresponds
to $H_{dI}$, $P_{i2}$ to $H_{uI}$ and $P_{i3}$ to $S_I$, respectively.

The first term in Eq.\,(\ref{eq:coupling}) implies that
$\sin\theta_{U,V} \simeq 1$ (yielding a Higgsino-like $\chi_1^\pm$),
$P_{13}\simeq 1$ and that larger
values of $\lambda$ are needed in order to 
enhance $C^{\rm eff}_{a_1} (\gamma\gamma)$ for the singlet-like $a_1$. 
On the other hand, for the doublet-like pseudoscalar, $a_2$, an
enhancement in $C^{\rm eff}_{a_2} (\gamma\gamma)$ requires either
$\cos\theta_U \sin\theta_V$ or $\sin\theta_U \cos\theta_V$ to be
non-negligible. This can be realized only in a very limited region of
the parameter space where $M_2 \simeq \mueff$ and not too large in
order to keep $m_{\chi^{\pm}_1}$ low. Moreover, in this case, the
mixing angles in the chargino sector read
\begin{equation}
\label{eq:thetaUV}
\theta_{U,V} \simeq
\arctan{\Bigg[\frac{\pm2M_W^2\frac{1-\tan^2\beta}{1+\tan^2\beta} -
    2\sqrt{(M_W^2+\mueff^2)^2 - \mueff^4}}{\sqrt{2}\,M_W\,\mueff (1+\tan\beta)}\Bigg]}\,,
\end{equation}
where $m_W$ is the mass of $W$ boson. The sign of the first term
implies that the enhancement can only be seen when $a_2$ has a leading
$H_{dI}$ component so that the term in Eq.\,(\ref{eq:thetaUV})
proportional to $ \sin\theta_U \cos\theta_V$ is dominant. Evidently,
in this case the $a_2b \bar{b}$ coupling, and in turn BR($a_2
\rightarrow b\bar{b}$), will also get enhanced. Consequently, a
contribution from $a_2$ will provide no significant excess in the
$\gamma\gamma$ signal rate, defined in Eq.\,(\ref{eq:Rxsct1}).

The above explanation also precludes such a scenario in the MSSM,
where the pseudoscalar, $A$, is doublet-like. Besides, as noted
in\cite{Christensen:2012ei,Carena:2013qia}, in the MSSM in order to
obtain the lightest CP-even Higgs boson, $h$, with mass around 125\gev,
$m_A$ is required to be $\gtrsim 300$\gev, which is
the so-called decoupling regime of the model. On the other hand, while
it is also possible to have a $\sim$125\gev\ $H$, the heavier CP-even
Higgs boson of the MSSM, this can only be achieved for $95\gev < m_A <
110\gev$, in a tiny portion of the `non-decoupling regime'. This
region is, moreover, disfavored by the constraints from flavor 
physics\cite{Scopel:2013bba,Boehm:2013qva}.     

In the fully constrained version of the NMSSM, unification of \akap\
and \azero\ at the GUT scale introduces tension between the masses of
$h_1$ and $a_1$, not allowing both to acquire values
$\lesssim 125$\gev\ simultaneously. There, in order to obtain the correct $h_1$ mass, large
negative values of \azero\ are necessary so that the mixing term
($\frac{X_t}{M_{\textrm{SUSY}}} \simeq \frac{A_t}{M_{\textrm{SUSY}}}$)
can be maximized. A light $a_1$, on the other hand requires
small \akap\ at \msusy, 
which in turn implies small \akap\ at the GUT scale, owing to the effects
of running. Moreover, small values of $\mueff$, necessary to obtain light
Higgsino-like charginos, additionally
limit the running of $A_t$ in the CNMSSM\cite{Kowalska:2012gs}.
Therefore, to obtain a SM-like $\sim$
125\gev\ $h_1$ and a pseudoscalar with a similar mass and a
non-negligible $\gamma\gamma$ rate 
one has to look beyond the MSSM and the CNMSSM, hence we analyse 
the CNMSSM-NUHM here. 

Through the mechanism explained above, a more precise measurement of the reduced effective
coupling, $C_{a_1}(\gamma\gamma) \equiv \frac{C^{\rm eff}_{a_1}(\gamma\gamma)}
{C^{\rm eff}_{h_{\rm SM}}(\gamma\gamma)}$, can yield an effective limit
on the mass of the lighter chargino through\footnote{Assuming a
  singlet-like $a_1$, which implies $P_{13} \simeq1$, and a Higgsino-like
  $\chi^\pm_1$ so that $\sin\theta_{U,V} \simeq 1$.} 
\begin{equation}
\label{eq:Ca1_limit}
C_{a_1}(\gamma\gamma) \simeq \lambda \times \frac{130\ \textrm{GeV}}{m_{\chi^{\pm}_1}}\,, 
\end{equation}
for $m_{a_1} \simeq 125$\gev. The bound obtained on the mass of
$\chi_1^\pm$ is also an effective upper limit on
the mass of the lightest neutralino, $\chi$ ($\equiv \chi_1^0$). 

Having described the mechanism for enhancing the $\gamma\gamma$ decay
rate of $a_1$, we now discuss the actual quantity
used for comparison with the experimentally observed $\gamma\gamma$
rate. In terms of the reduced effective couplings, $C_{a_1}(\gamma\gamma)$ and
$C_{a_1}(dd)$, of $a_1$ to
$\gamma\gamma$ and $b\bar{b}$, respectively, the signal rate, given in Eq.\,(\ref{eq:Rxsct1}),
can be rewritten for the $b\bar{b}h$ production mode as
\begin{equation}
R^{bb}_{\gamma\gamma} (a_1) = C^2_{a_1}(dd)\,C^2_{a_1}(\gamma\gamma)\,\frac{\Gamma^{\textrm{total}}_{h_\textrm{SM}}}{\Gamma^{\textrm{total}}_{a_1}} \simeq |P_{11}''|^2\,\lambda^2\,\Big(\frac{130\ \textrm{GeV}}{m_{\chi^{\pm}_1}}\Big)^2\,\Big(\frac{1}{\Gamma^{\textrm{total}}_{a_1} / \Gamma^{\textrm{total}}_{h_\textrm{SM}}}\Big)\,,
\label{eq:rcs_general}
\end{equation}
where $|P_{11}''| \simeq \big|\frac{\lambda(A^{\rm SUSY}_{\lambda}-2\kappa
  s)v}{\mu(A^{\rm SUSY}_{\lambda}+\kappa s)}\big|$ and
$\Gamma^{\textrm{total}}_{a_1}$ and $\Gamma^{\textrm{total}}_{h_\textrm{SM}}$ denote the theoretical values of
the total widths of $a_1$ and a SM Higgs boson with the same mass as
$a_1$, respectively. 
The dependence of the above expression on
$\tan\beta$ is not straightforward, since it only enters indirectly through 
$\Gamma^{\textrm{total}}_{h_\textrm{SM}}/\Gamma^{\textrm{total}}_{a_1}$. 
Eq.\,(\ref{eq:rcs_general}) also shows that, as noted in the
Introduction, the conditions necessary to enhance
$C_{a_1}(\gamma\gamma)$, i.e., large $\lambda$ and small $\mu$, also yield
an enhanced $|C_{a_1}(dd)| \simeq |P_{11}''|$.

In \refsec{Model} we will
use Eqs.\,(\ref{eq:Ca1_limit}) and (\ref{eq:rcs_general}) to obtain
an effective upper limit on $m_{\chi_1^\pm}$ and the mass of $\chi$, $m_\chi$, in our model
under consideration.

\subsection{$b\bar{b}/\tau^+\tau^-$ decay of the pseudoscalar}
\label{sec:a1tautau}

The signal rate in these decay modes can be written, following
Eq.\,(\ref{eq:rcs_general}), as
\begin{equation}
R^{bb}_{b\bar{b}/\tau^+\tau^-} (a_1) \simeq \frac{|P_{11}''|^4}{\Gamma^{\textrm{total}}_{a_1} / \Gamma^{\textrm{total}}_{h_\textrm{SM}}}\,,
\label{eq:rcs_tau}
\end{equation}
It should be noted in the above expression that both the $b\bar{b}$ and $\tau^+\tau^-$
decay rates scale with the same reduced coupling $C_{a_1}(dd)$. Both
these decay channels, therefore, show exactly the same behavior as far as
their signal rates are concerned, despite the fact that
BR($a_1\rightarrow\tau^+\tau^-$) is considerably smaller than
BR($a_1\rightarrow b\bar{b}$). From an experimental point of view, the $b\bar{b}$
decay mode will result in 4 $b$-jets which may be quite challenging to tag
owing to the large hadronic background, although this mode has been
visited in the past\cite{CMS-H-twiki}. The $\tau^+\tau^-$ decay mode,
on the other hand, is subject to a much smaller leptonic background and
is in fact the preferred mode for analysing possibly supersymmetric
Higgs bosons.
 

\section{\label{Model}The CNMSSM-NUHM}

In the fully constrained NMSSM universality conditions are imposed on
the dimensionful parameters at the GUT scale. This leads to a unified
gaugino mass parameter, \mhalf, besides \mzero\ and \azero, with
\alam\ and \akap\ also unified to the latter.  Thus, given the correct
value of the mass of $Z$ boson, $m_Z$, \mzero, \mhalf, \azero\ and
\lam, taken as an input parameter at \msusy, constitute the only free
parameters in the CNMSSM.

In the partially unconstrained version of the model, the CNMSSM-NUHM,
the soft masses of the Higgs fields, $m_{H_u}$, $m_{H_d}$ and $m_S$, as
well as the soft trilinear coupling parameters \alam\ and \akap\ are
taken as free parameters at the GUT scale, instead of assuming their unification with \mzero\
and \azero, respectively. Through the minimization
conditions of the Higgs potential the three 
mass parameters $m_{H_u}$, $m_{H_d}$ and $m_S$ at the electroweak scale can be
traded for the parameters \kap, \mueff\ and \tanb. The model is thus defined in terms of
the following eight continuous input parameters: 

\begin{center}
\mzero, \mhalf, \azero, \tanb, \lam, \kap, \mueff, $\alam=\akap$.
\end{center}

The unification of \alam\ and \akap\ at the GUT scale assumed above
is in general not necessary in the CNMSSM-NUHM. In fact, one can
argue that the restriction on \azero\ for $\mueff<0$ and the resultant
tension between $m_{h_1}$ and $m_{a_1}$ discussed in the
previous section can be relaxed by not imposing such a
condition. In that case, the effect of large \alam\ can be counter-balanced
by increasing \akap\ independently, thus still yielding physical $a_1$
solutions. However, this unification condition has minimal impact on the allowed
parameter space of the model for our purpose,   
since, as we shall see later, we can still exploit the
interesting phenomenology of the model while keeping the number of
free parameters to a minimum. This is also consistent with the fully
constrained version of the model that we studied
earlier\cite{Kowalska:2012gs}, where \akap\ and \alam\ were set equal
to \azero\ at the GUT scale, even though $m_S \neq m_0$.



\section{\label{Results}Methodology and results}

We perform scans of the parameter space of CNMSSM-NUHM
requiring both $h_1$ and $a_1$ to have masses near 125\gev. 
We impose the latest 95\% confidence level (CL) exclusion limit on the
(\mzero,\,\mhalf) space of mSUGRA/CMSSM obtained by the ATLAS
collaboration from
two same-sign leptons and jets in the final state at
$\sqrt{s}=8$\tev\ with 20.7/fb of data\cite{ATLAS-CONF-2013-007}.
 It has been verified in\cite{Kowalska:2012gs,Kowalska:2013hha}
  that such a limit, obtained originally for the CMSSM, generally has negligible dependence on the Higgs sector parameters
  and is, therefore, applicable to any $R$-parity conserving SUSY model
  with unified \mzero\ and \mhalf. 
We also impose the lower limit, $m_{\chi^{\pm}_1} >
94$\gev\cite{Beringer:1900zz}, on the lightest chargino mass in our scans. Furthermore,
we include Gaussian likelihoods for the most significant $b$-physics
observables, with their measured mean values and
errors taken as: 
\begin{itemize}
\item $\brbsmumu = (3.2^{+1.5}_{-1.2}\pm 0.32) \times 10^{-9}$,
\item $\brbutaunu = (1.66\pm 0.66 \pm 0.38) \times 10^{-4}$,
\item $\brbxsgamma = (3.43\pm 0.22 \pm0.21) \times 10^{-4}$ and
\item $\delmbs = (17.72\pm0.04\pm2.4$)\,ps$^{-1}$.
\end{itemize}

For testing the compatibility of the regions of interest against the
dark matter direct detection cross section, \sigsip, we use the
XENON100 90\% CL exclusion limits\cite{Aprile:2012nq}. Note that we
neglect the \amu\ constraint here since it is well known that the
regions where correct \amu\ can be obtained in the parameter spaces of
SUSY models with unification of squark and slepton soft masses are
strongly disfavored by the direct SUSY searches at the
LHC\cite{Roszkowski:2012uf,Kowalska:2012gs,Kowalska:2013hha}. However,
in order to minimize the deviation from experimental value of \amu\
and also to release the tension between $m_{h_1}$ and $m_{a_1}$, as
discussed at the end of \refsec{sec:a1mass}, we shall use $\mueff>0$,
unless stated othewise. We also note here that no likelihood function
was implemented in our scans for the relic density constraint. However, in
all our results below we only show points with neutralino relic density,
$\Omega_\chi h^2$, lying in the $\pm 2\sigma$ range, $0.087 <
\Omega_\chi h^2 < 0.137$, around the central experimental value
(again, unless stated otherwise), after taking into 
account $10\%$ error on the theoretical calculation. We use a slightly extended range of
the allowed Higgs boson mass, $122\ \textrm{GeV} < m_{h_1,a_1} < 130$\gev,
compared to the mass measurements of the observed boson at the LHC in
order to take into account large theoretical and experimental errors.
Finally, for all the points considered, $h_1$ is always SM-like, with
$R^{bb}_{X/V} (h_1)\simeq 1$. 

The numerical analysis was performed using the BayesFITS package
which engages several external, publicly available tools: 
MultiNest\cite{Feroz:2008xx} for sampling of the CNMSSM-NUHM parameter space;
\nmssmtools\ v3.2.4\cite{NMSSMTools} for computing SUSY mass spectrum,
Higgs BRs and reduced couplings, as well as \delmbs\ for a given NMSSM point; \superiso\
v3.3\cite{superiso} for calculating \brbxsgamma, \brbsmumu\ and \brbutaunu. DM observables such as the relic density
and \sigsip\ are calculated with \micromegas\
v2.4.5\cite{micromegas}. 

\subsection{$\gamma\gamma$ rate enhancement}

\begin{figure}[ht!]
\centering
\subfloat[]{%
\label{fig:-a}%
\includegraphics*[height=7cm]{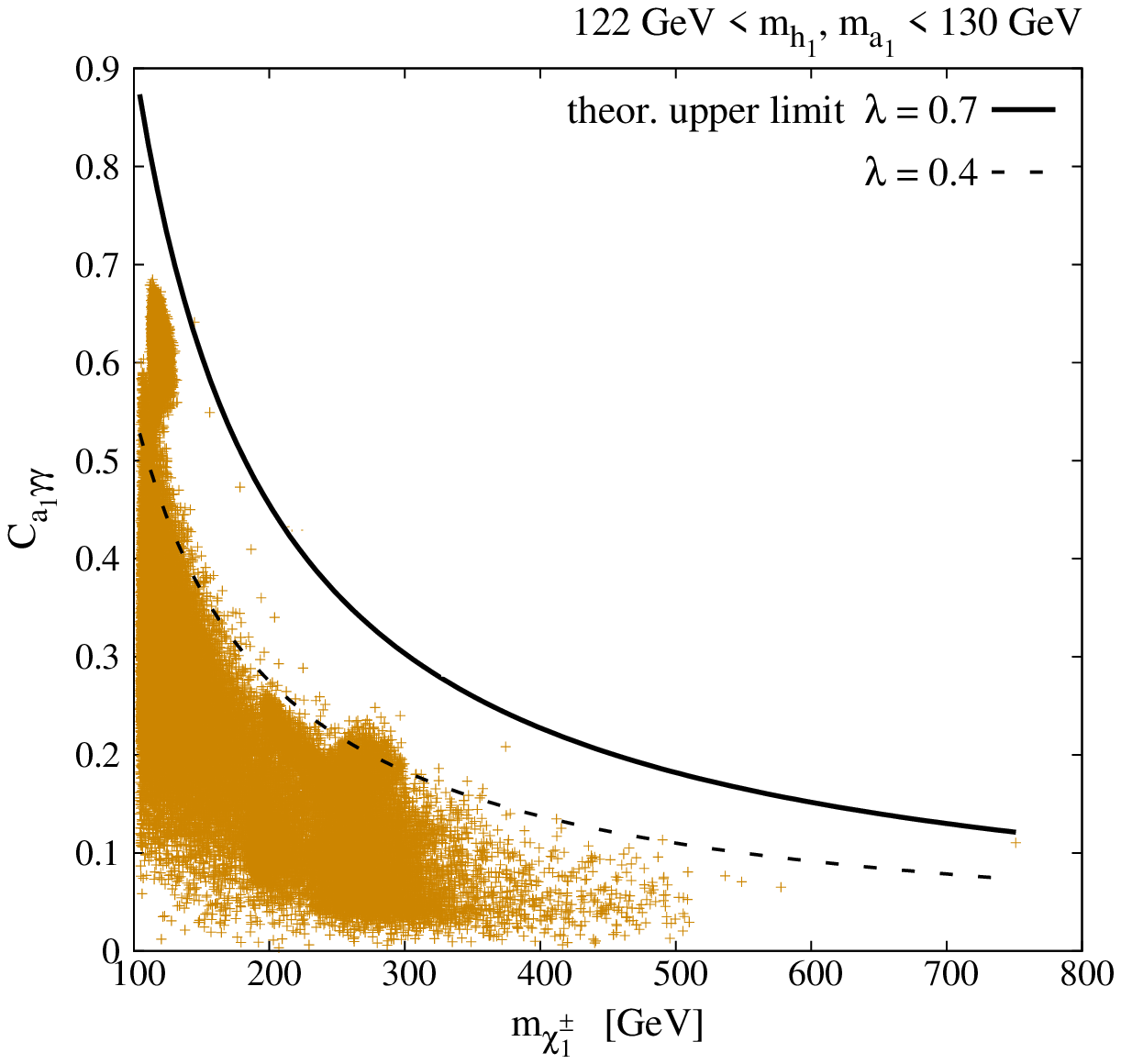}
}%
\hspace{0.5cm}%
\subfloat[]{%
\label{fig:-b}%
\includegraphics*[height=7cm]{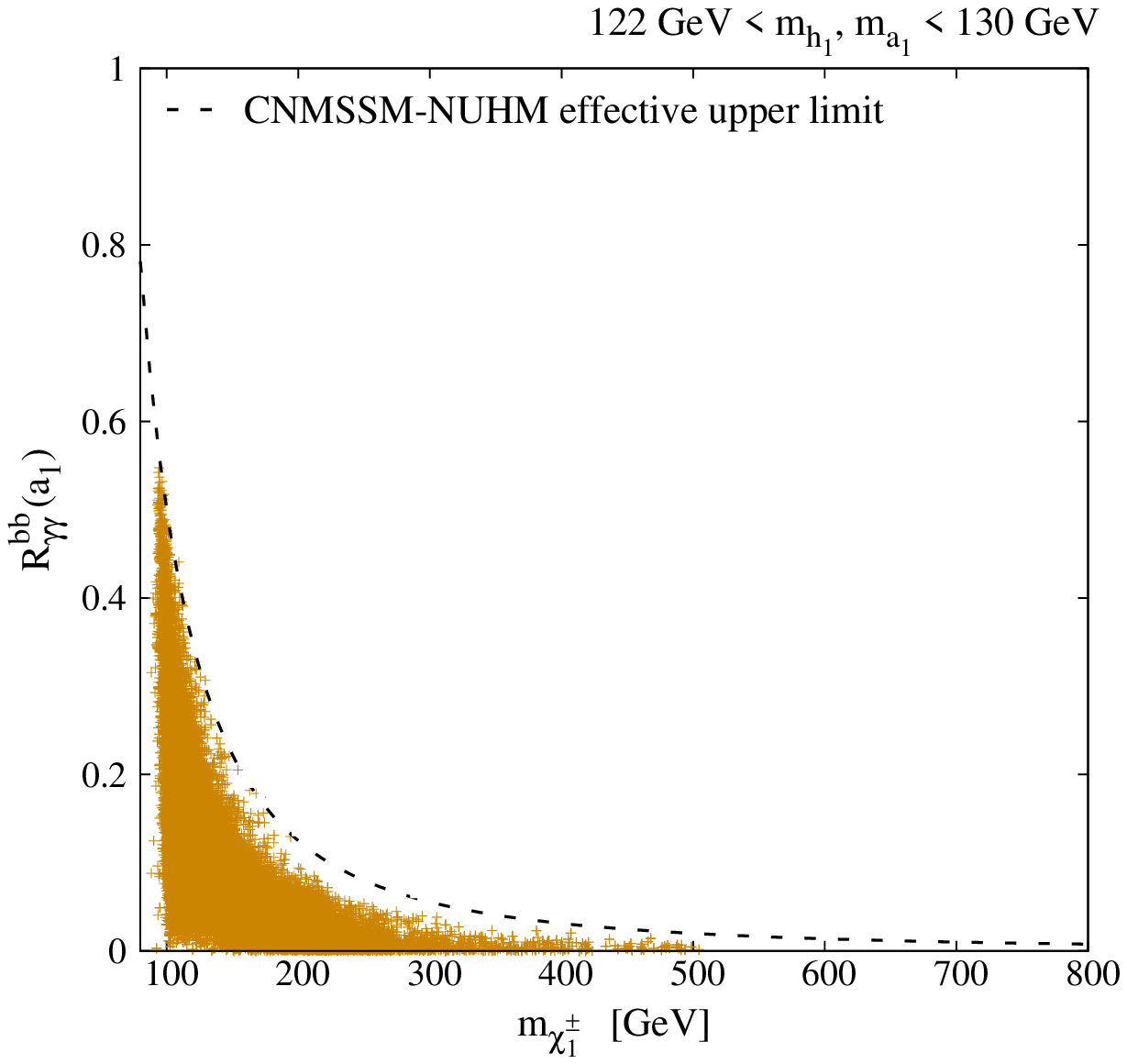}
}%
\caption[]{(a) Distribution of points obtained in our scan of the
  CNMSSM-NUHM parameter space
 in the ($m_{\chi_1^\pm}$,\,$C_{a_1}(\gamma\gamma)$) plane. The dashed
 line shows the effective upper limit observed in the scan. The solid
 line is based on a perturbative upper limit on
 $\lambda$ and is shown for comparison. (b)
Distribution of points in the ($m_{\chi_1^\pm}$,\,$R^{bb}_{\gamma\gamma}(a_1)$) 
plane. The dashed line shows the effective upper limit observed in our scan.}
\label{fig:cplg}
\end{figure}

As noted in \refsec{Pseudo}, the scenario under consideration
requires low values of \mueff\ giving a light Higgsino-like
$\chi_1^\pm$ and correspondingly a $\chi$ with significant Higgsino
component. Under these conditions the upper limit on $m_{\chi_1^\pm}$ and
$m_\chi$ can be obtained in the CNMSSM-NUHM from
Fig.\,\ref{fig:cplg}, where $C_{a_1}(\gamma\gamma)$ and
$R^{bb}_{\gamma\gamma}(a_1)$ are shown as functions of
$m_{\chi_1^{\pm}}$ in (a) and (b), respectively. For all points in the
plots we assume $122\ \textrm{GeV} < m_{h_1,a_1} < 130$\gev.

The parameter space of the CNMSSM-NUHM giving an enhancement in the
$\gamma\gamma$ rate due to a $\sim$125\gev\ $a_1$ can 
in fact be divided into three main regions depending on the composition of
$\chi$:\\
i) the singlino-Higgsino region,\\ 
ii) the Higgsino region, and\\ 
ii) the focus-point (FP) region. \\
Below we discuss the results for each of these regions separately. 

\subsubsection{Singlino-Higgsino region}

\begin{figure}[p]
\centering
\subfloat[]{%
\label{fig:-a}%
\includegraphics*[height=6.5cm]{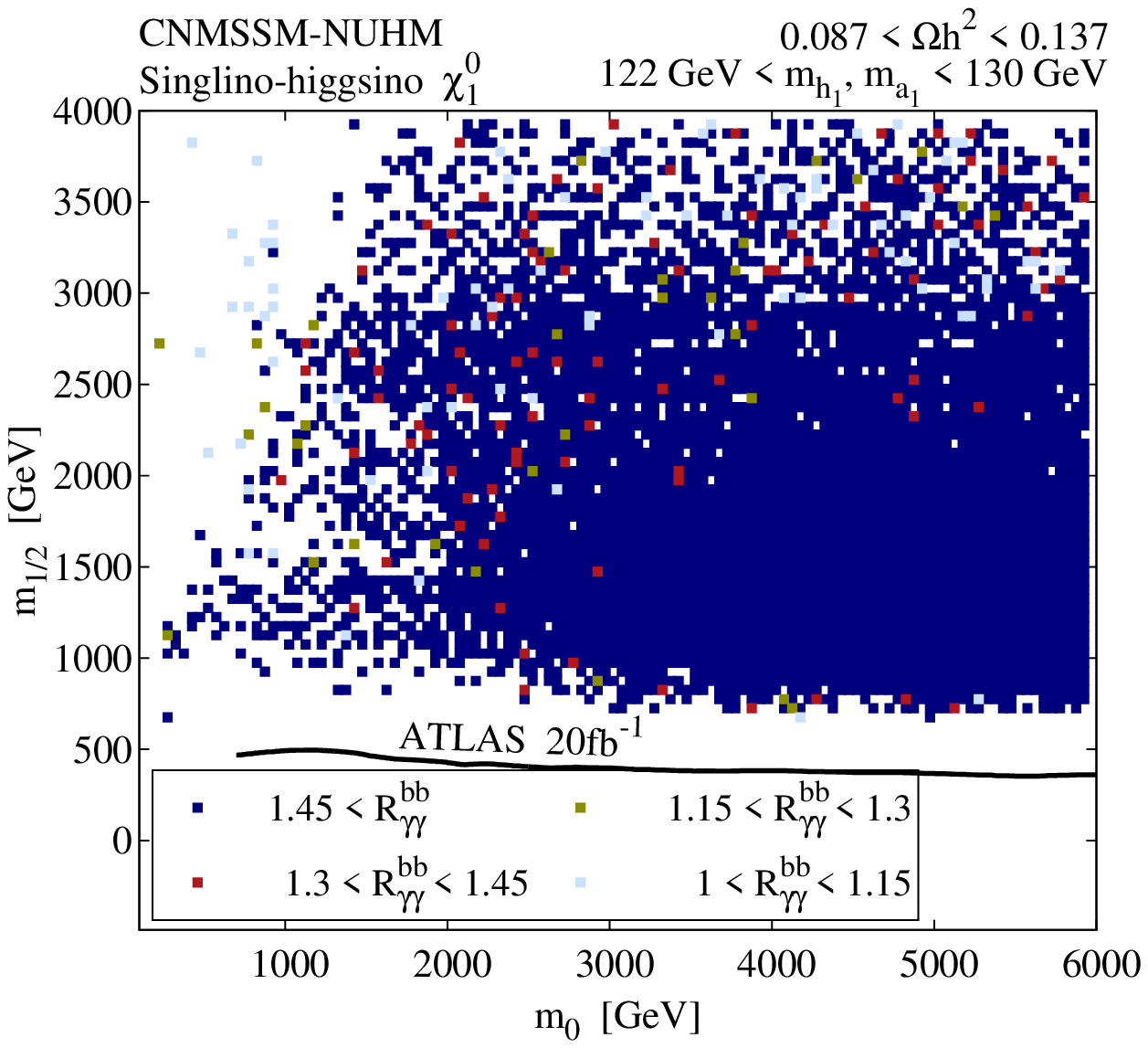}
}%
\hspace{0.5cm}%
\subfloat[]{%
\label{fig:-b}%
\includegraphics*[height=6.5cm]{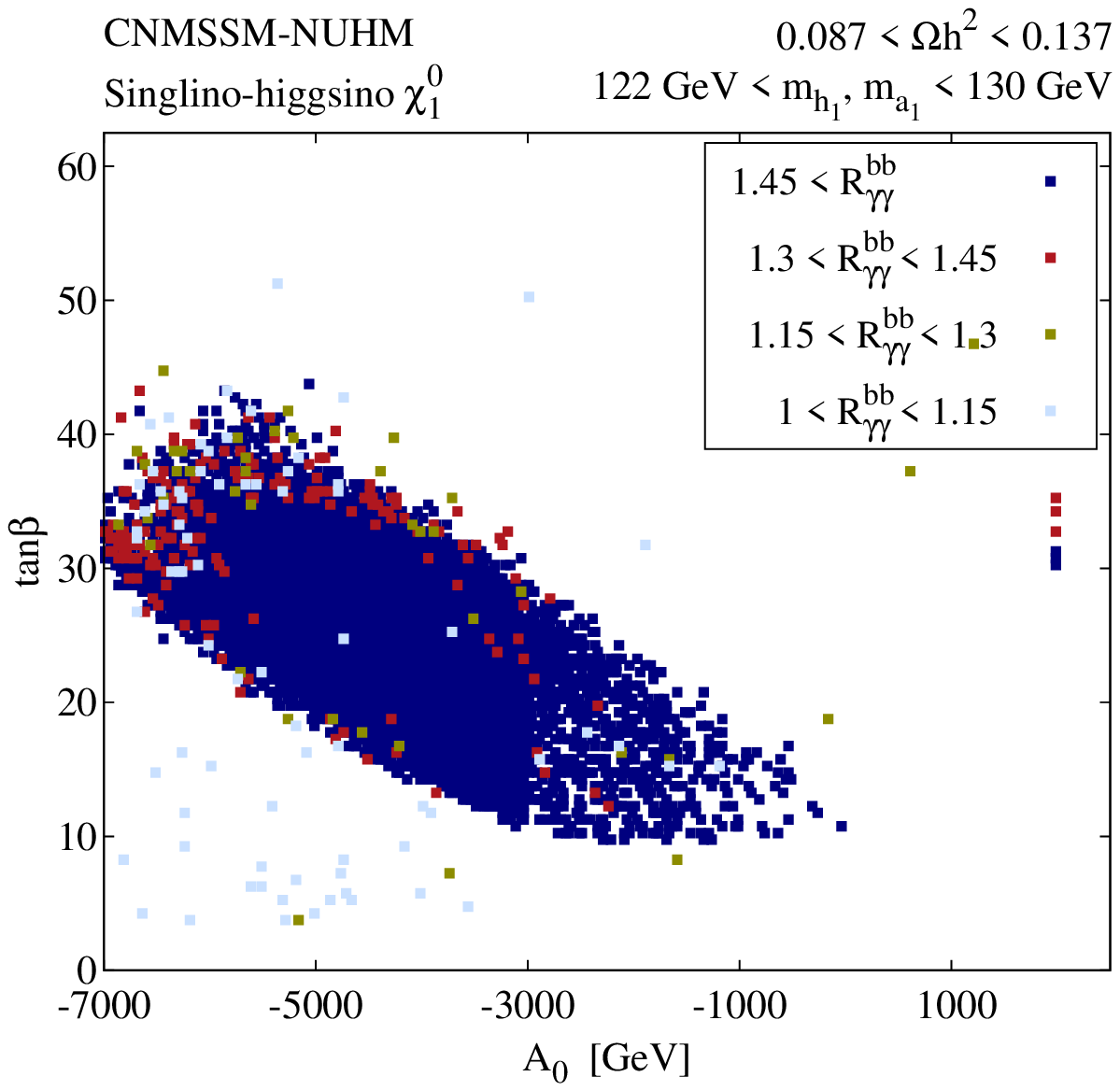}
}%

\subfloat[]{%
\label{fig:-a}%
\includegraphics*[height=6.5cm]{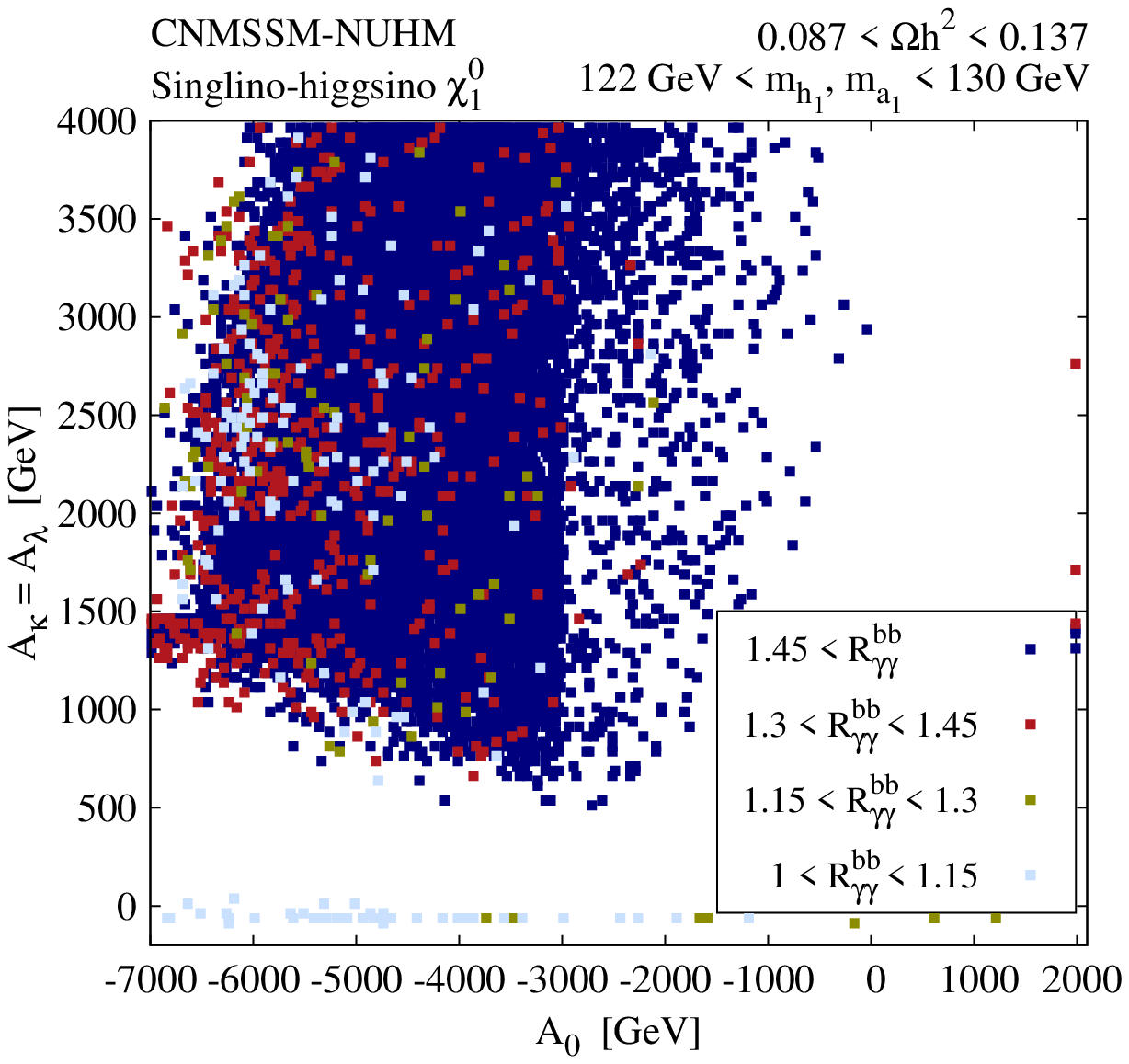}
}%
\hspace{0.5cm}%
\subfloat[]{%
\label{fig:-b}%
\includegraphics*[height=6.5cm]{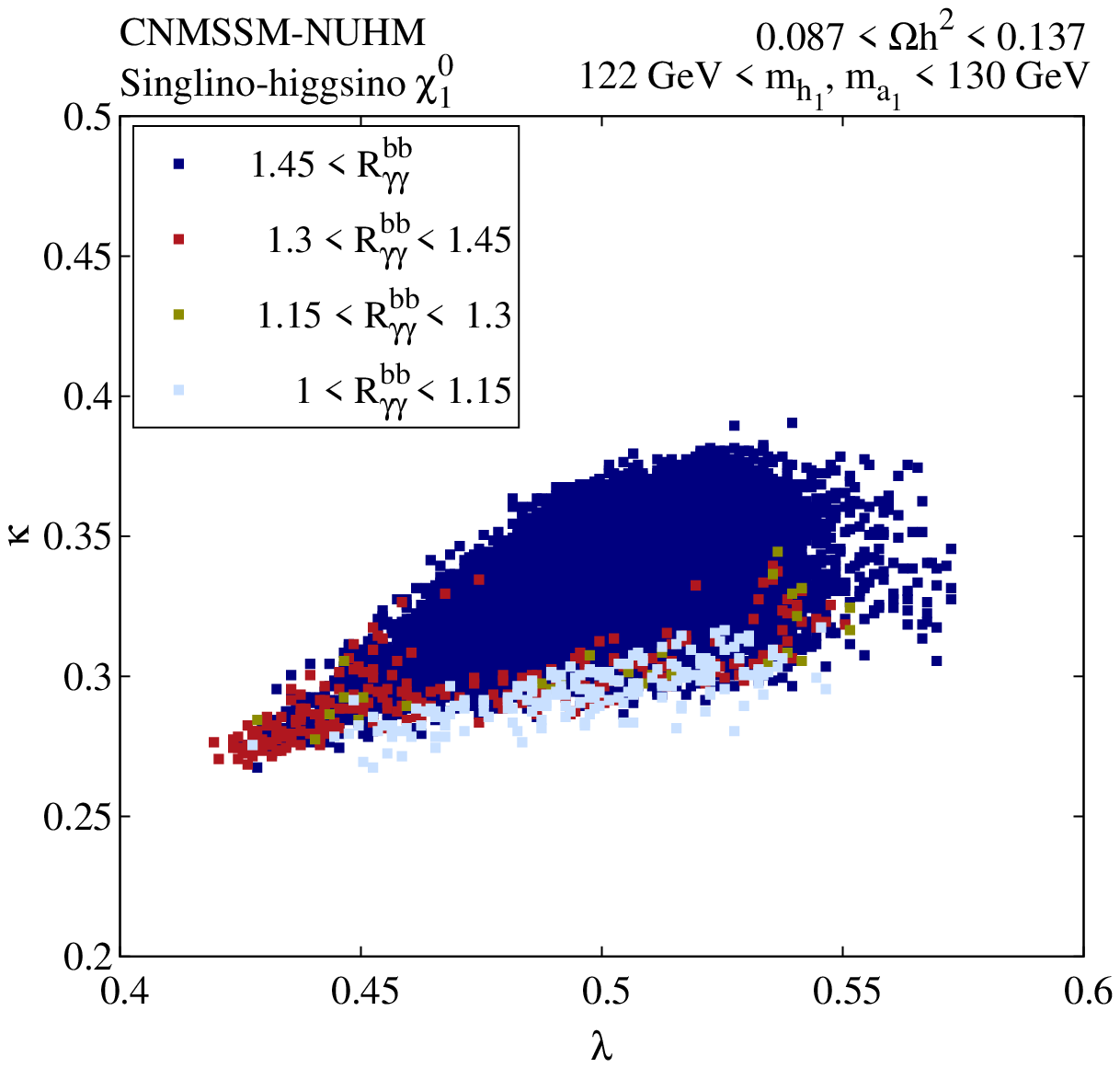}
}%

\subfloat[]{%
\label{fig:-a}%
\includegraphics*[height=6.5cm]{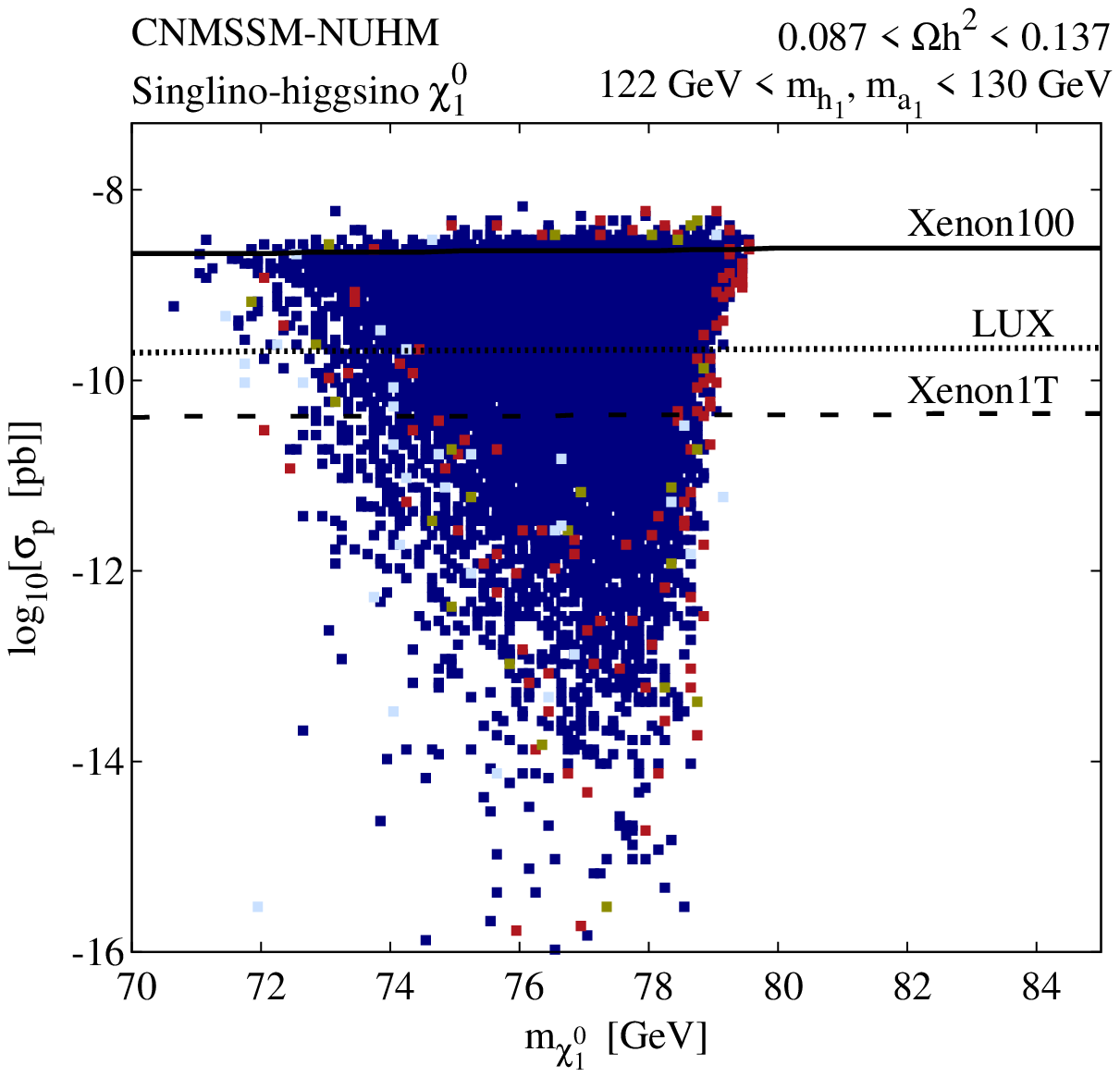}
}%
\hspace{0.5cm}%
\subfloat[]{%
\label{fig:-b}%
\includegraphics*[height=6.5cm]{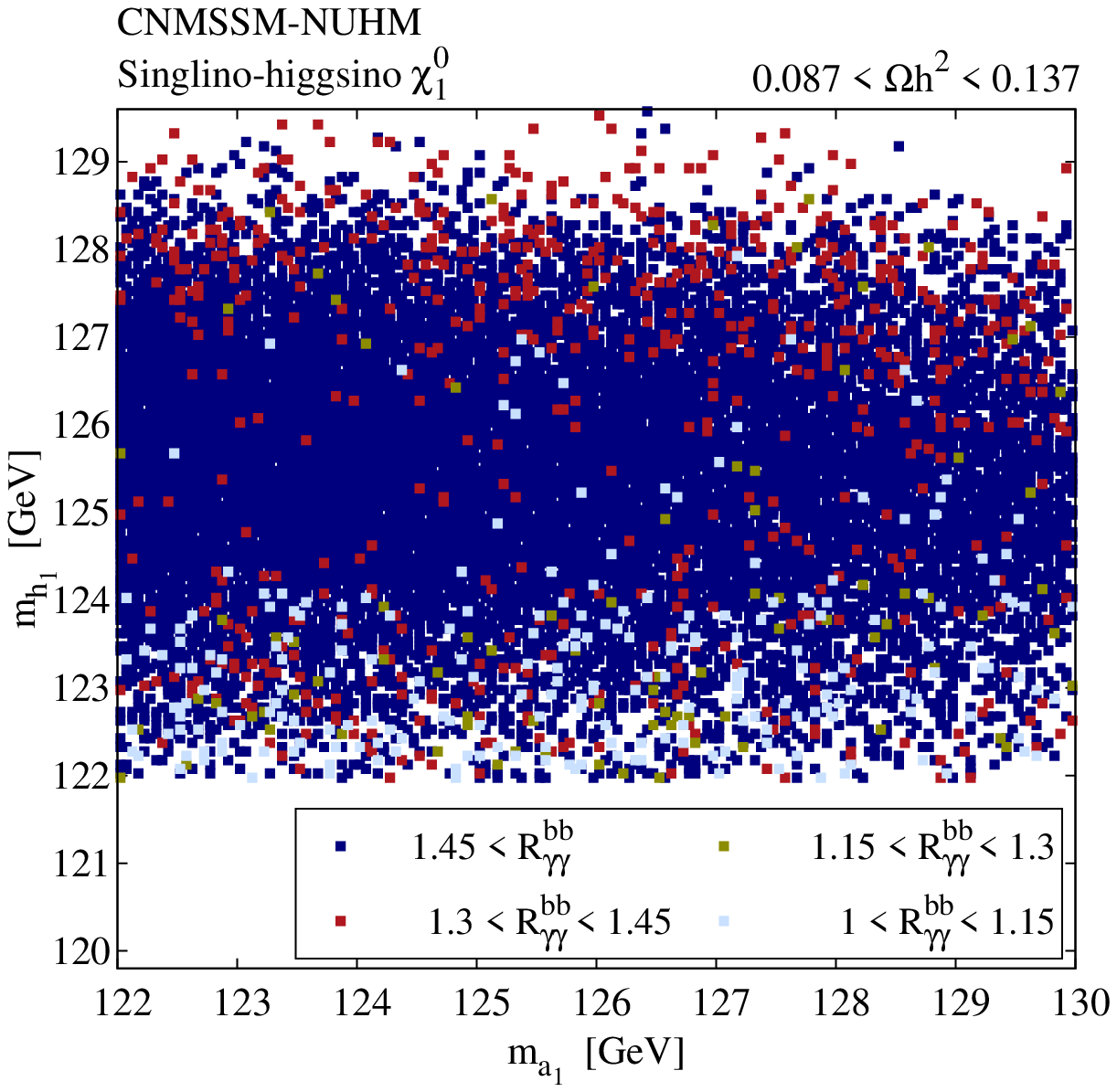}
}%
\caption[]{(a)-(d) Ranges of CNMSSM-NUHM paramaters corresponding to
  the singlino-Higgsino region. (e) \sigsip\ obtained for this region as a function
  of $m_\chi$. (f) Ranges of $m_{h_1}$ and $m_{a_1}$ obtained in this
  region. See text for details.}
\label{fig:SIN}
\end{figure}

This region is defined by $\chi$ being a mixture of a large Higgsino
component and a smaller but important singlino component. Owing to the
significant singlino component (20\%--30\%) the neutralino will
interact very weakly with matter and will thus result in large relic
abundance unless it has a small mass and consequently large
annihilation cross-section. In Fig.\,\ref{fig:SIN}a we see the
distribution of this region in the (\mzero,\,\mhalf) plane. Light blue
squares correspond to points with $1 < R_{\gamma\gamma}^{bb} (h_1 +
a_1) \leq 1.15$, green squares to points with $1.15 <
R_{\gamma\gamma}^{bb} (h_1 + a_1) \leq 1.3$, red squares to points
with $1.3 < R_{\gamma\gamma}^{bb} (h_1 + a_1) \leq 1.45$ and green
squares to points with $R_{\gamma\gamma}^{bb} (h_1 + a_1) > 1.45$.
Also shown in the figure is the current 95\%
CL exclusion limit from ATLAS obtained with 20/fb of data.

While \mzero\ is widely distributed, intermediate-to-large values of \mhalf\ are
favored for allowing a neutralino with a negligible bino
component for small positive \mueff. In Fig.\,\ref{fig:SIN}b the
favored ranges of $\tan\beta$ and \azero\
parameters are shown. We see that the enhancement in 
$R_{\gamma\gamma}^{bb} (h_1 + a_1)$ decreases as $\tan\beta$
increases. The reason for this is as follows. The 
enhancement in $R_{\gamma\gamma}^{bb} (a_1)$ grows with $\lambda$,
according to Eq.\,(\ref{eq:rcs_general}). However, large values of
$\lambda$ can only give correct $m_{h_1}$ for not too large values of
$\tan\beta$. This is because larger values of $\tan\beta$ result in
an enhanced Yukawa coupling of $h_1$ to $b\bar{b}$ and
$\tau\bar{\tau}$. This will make \alam\ run upwards faster from its GUT
scale value, which in
turn causes \akap\ to run downwards to larger negative values. That
will result in a decrease in $m_{h_1}$, since it has a significant singlet
component, while $m_{a_1}$ increases. \azero\ almost always takes large negative
values, in order to maximize $m_{h_1}$. We, therefore, hardly see any
points corresponding to positive \azero. 

In Fig.\,\ref{fig:SIN}c we show the distribution of points in the
(\azero,\,\akap) plane.  The interdependence of \azero\ and $\alam
=\akap$ is further illustrated by this figure.  \akap\ can take quite
large positive values at the GUT scale, even though it ought to be
negative at \msusy.  The reason is that, for large negative \azero\
values, \alam\, which is also positive, makes \akap\ run downwards.
In Fig.\,\ref{fig:SIN}d the ($\lambda$,\,$\kappa$) plane is shown.
The allowed range of $\lambda$ for a given region is subject to a
three-way tension. Large \lam\ is favored in order to obtain an
enhancement in the coupling of $a_1$ to $\chi_1^\pm$ but the condition
to obtain a SM-like $h_1$, on the other hand, prefers smaller
values. The small-to-intermediate range of \lam\ seen in the figure is
then a result of the compromise between these two conditions and,
additionally, of the requirement to achieve the desired $m_{a_1}$ by
generating $s$ ($=\mueff / \lam$) of the correct size. We note also in
the the figure small-to-intermediate values of $\kappa$ which are
required to maximize the singlino component of $\chi$, as the $5\times
5$ term in the neutralino mass matrix is equal to $2\kappa s$ ($\equiv
2\kappa \mueff /\lambda$).  Hence the smallness in $\kappa$ has to be
compensated by large values of \akap, as noted in
Fig.\,\ref{fig:SIN}c, for obtaining the correct $m_{a_1}$. Finally,
over the entire allowed ranges of $\lambda$ and $\kappa$ a large
enhancement in $R^{bb}_{\gamma\gamma} (h_1+a_1)$ is observed, although
this region corresponds to more fine-tuned values of these two
parameters compared to the other two regions, as we shall see later.

In Fig.\,\ref{fig:SIN}e we show the ($m_{\chi}$,\,\sigsip) plane for
this region. Also shown in
the figure are the actual 90\% CL exclusion limits from XENON100 as
well as the 90\% CL limits expected from
the LUX\cite{Akerib:2012ys} and XENON1T\cite{Aprile:2012zx}
experiments. A large number of points satisfying the XENON100 limit
lies below the projected 90\% CL XENON1T limit.  Note also that since
very small $m_{\chi}$ and consequently $m_{\chi_1^\pm}$ is favored by
this region almost all the points below the XENON100 line have a
highly enhanced $R^{bb}_{\gamma\gamma} (a_1)$, since $\chi_1^\pm$ also
appears in the denominator of Eq.\,(\ref{eq:rcs_general}). This is
also the reason why such points are achievable even with relatively
small values of $\tan\beta$, as seen in Fig.\,\ref{fig:SIN}b
earlier. This region yields the maximum enhancement, up to $\sim$60\%
or so, in $R^{bb}_{\gamma\gamma} (h_1+a_1)$ out of the three regions
discussed here and is, therefore, the most favorable of all.

In Fig.\,\ref{fig:SIN}f we show the distribution of $m_{h_1}$ versus
that of $m_{a_1}$. We note that this region can have fairly large
$m_{h_1}$, which is due to the combined effects of large negative
\azero\ as well as larger allowed values of $\lambda$.  Another
important feature of this region is that $h_2$ can also be almost mass
degenerate with $h_1$ and $a_1$, implying in that case a `triple
degeneracy' among the Higgses.  Again, while mass degenerate $h_1$ and
$h_2$ can explain the enhanced $\gamma\gamma$ rate in the $ggh$
production mode in the ATLAS data, in order to test the additional
degeneracy with $a_1$ one will have to explore the associated
production mode of Higgs with $b\bar{b}$. In the $b\bar{b}h$
production mode, such $h_2$ can further contribute $\sim$20\% of the
measured $\gamma\gamma$ rate.  Finally, \brbsmumu\ in this region
varies between $3\times 10^{-9}$ and $3.8\times 10^{-9}$ while
\brbxsgamma\ lies in the $2.8\times 10^{-4}$ to $3.3\times 10^{-4}$
range.

\subsubsection{Higgsino region}

\begin{figure}[p]
\centering
\subfloat[]{%
\label{fig:-a}%
\includegraphics*[height=6.5cm]{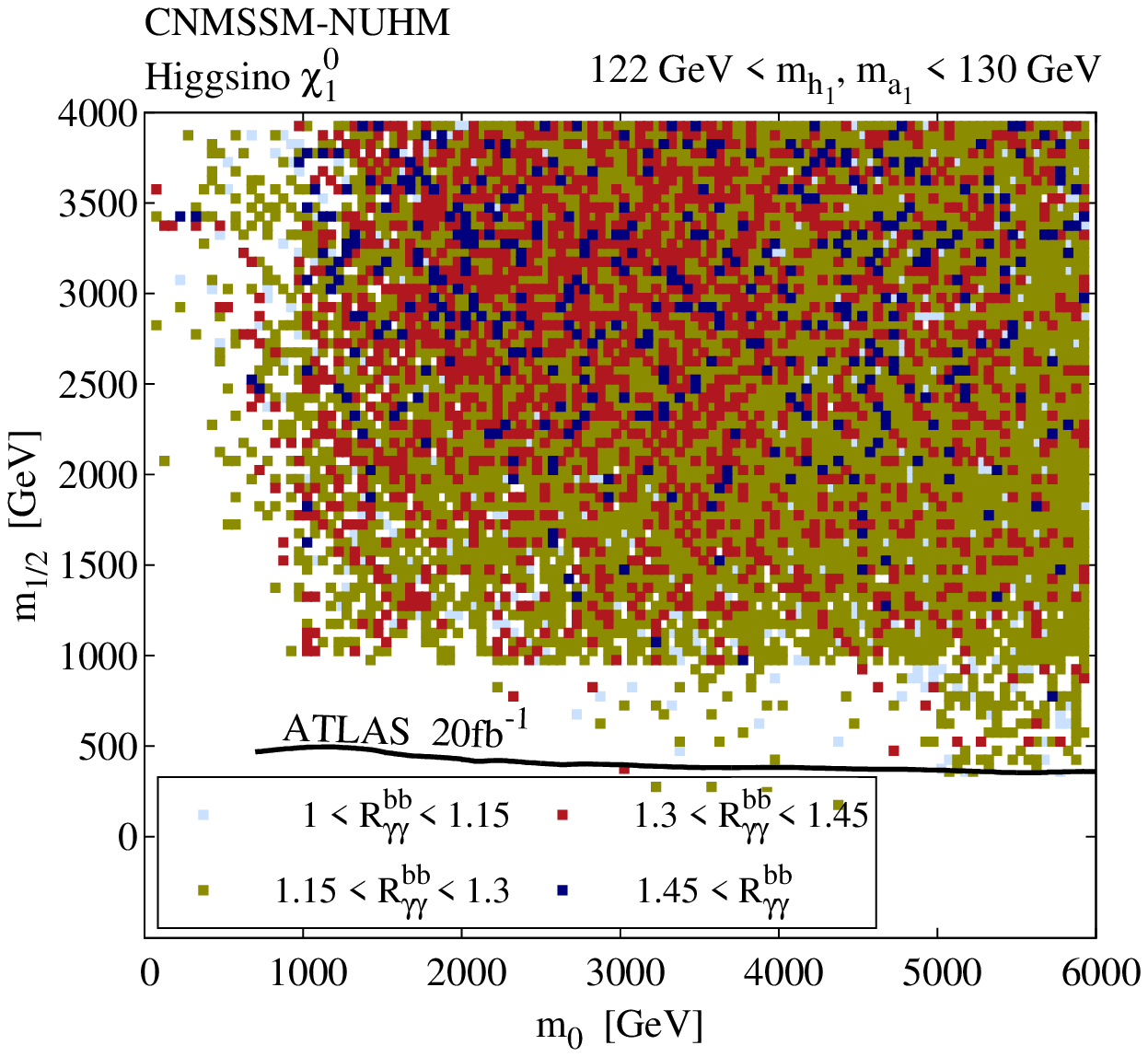}
}%
\hspace{0.5cm}%
\subfloat[]{%
\label{fig:-b}%
\includegraphics*[height=6.5cm]{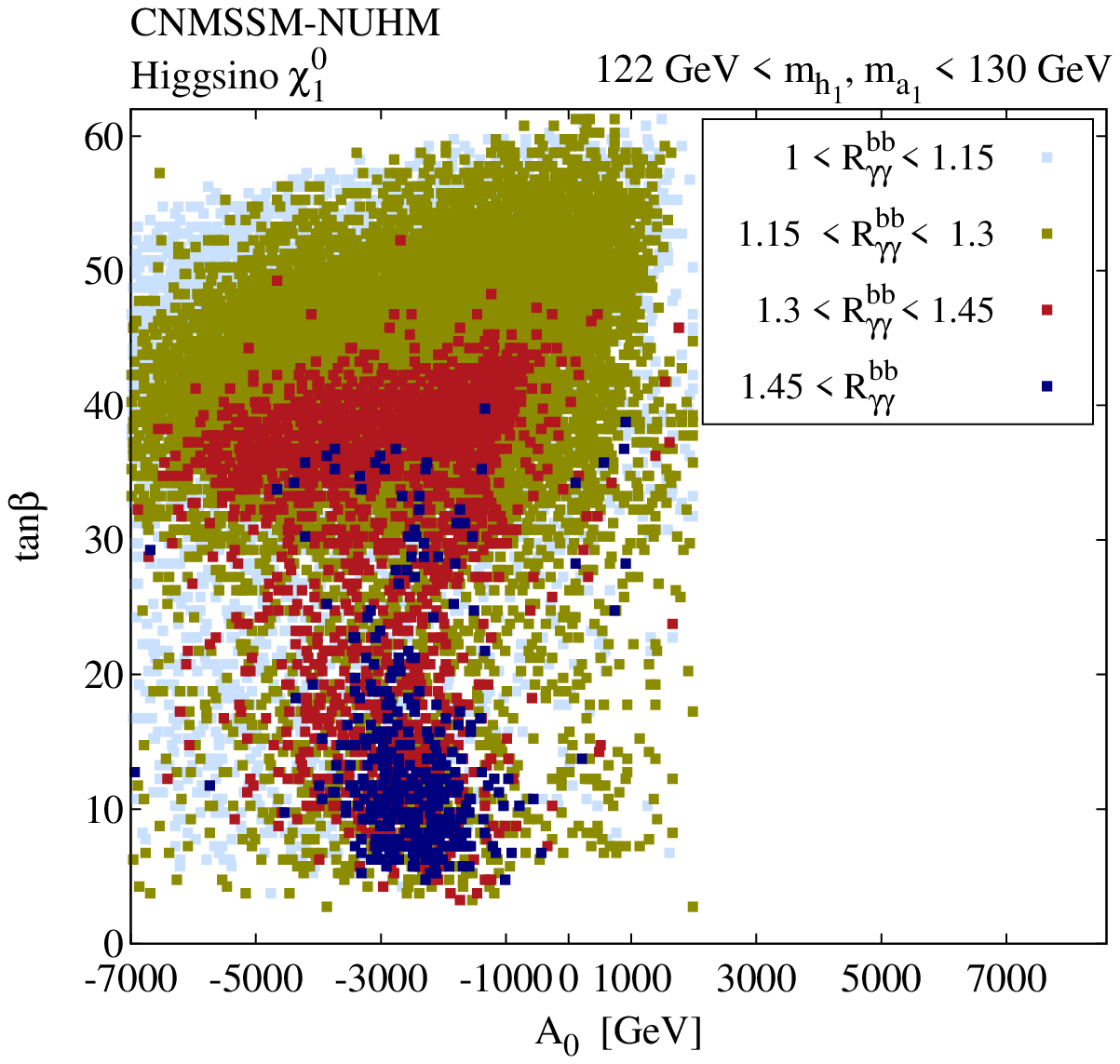}
}%

\subfloat[]{%
\label{fig:-a}%
\includegraphics*[height=6.5cm]{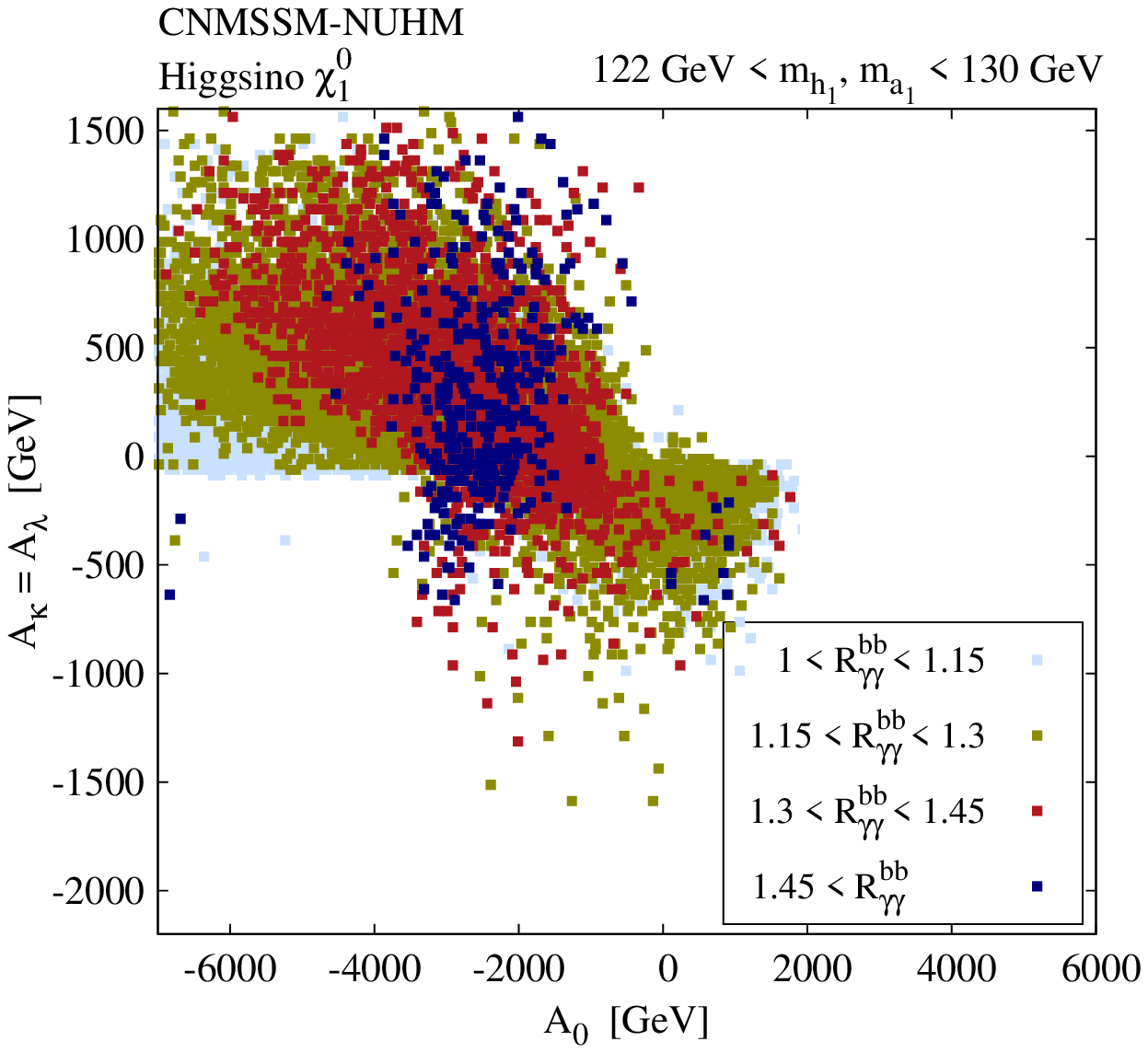}
}%
\hspace{0.5cm}%
\subfloat[]{%
\label{fig:-b}%
\includegraphics*[height=6.5cm]{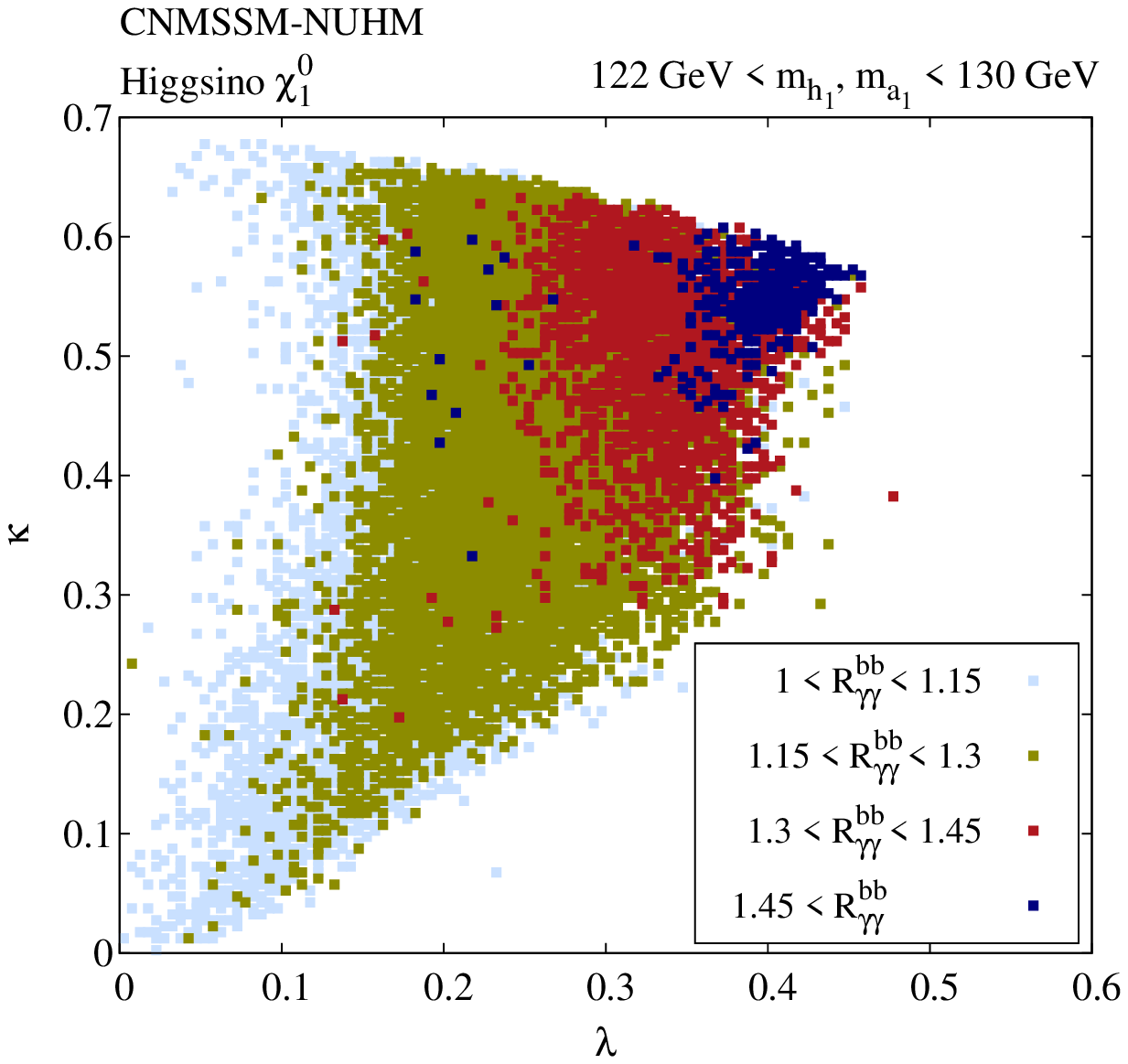}
}%

\subfloat[]{%
\label{fig:-a}%
\includegraphics*[height=6.5cm]{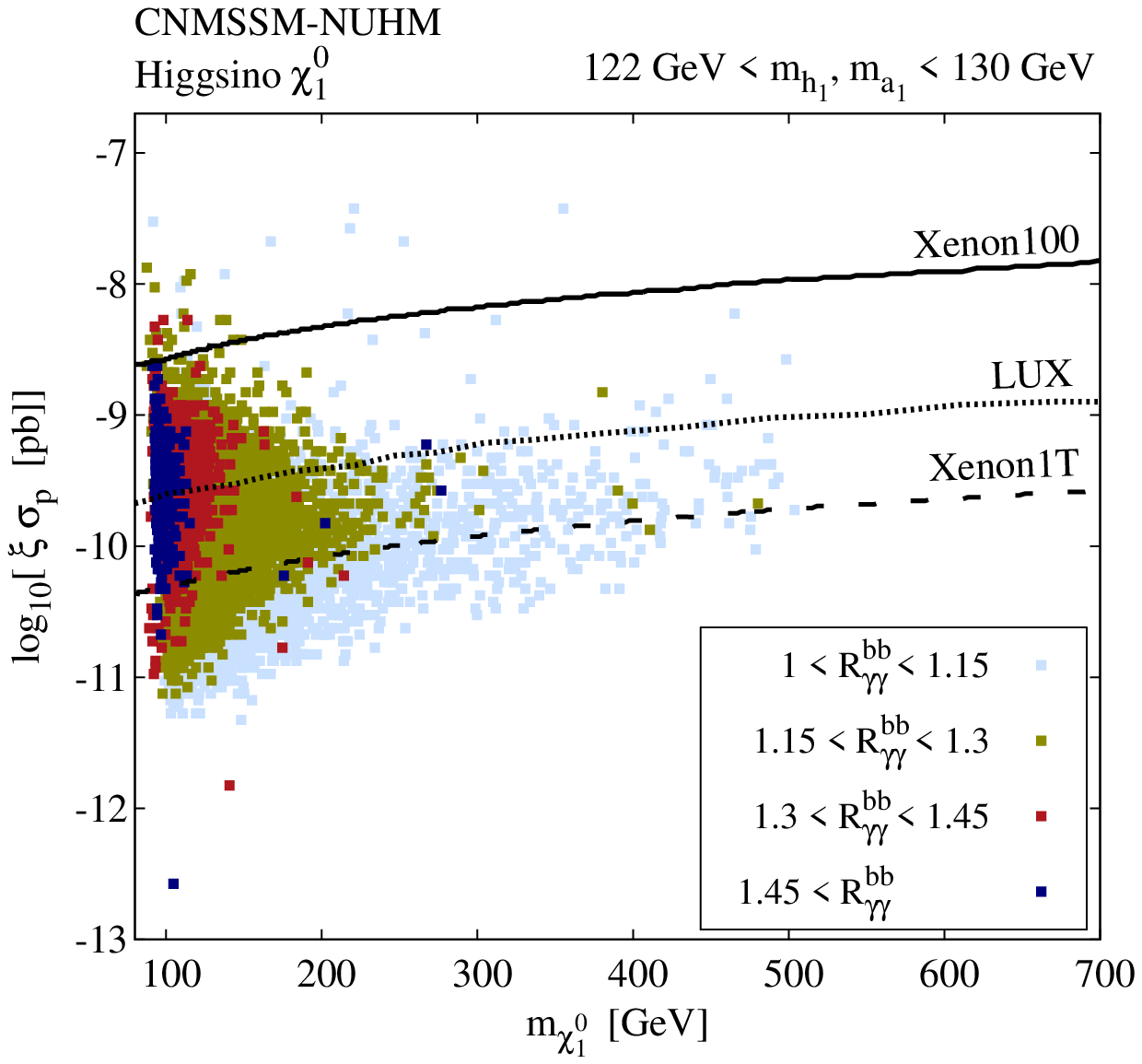}
}%
\hspace{0.5cm}%
\subfloat[]{%
\label{fig:-b}%
\includegraphics*[height=6.5cm]{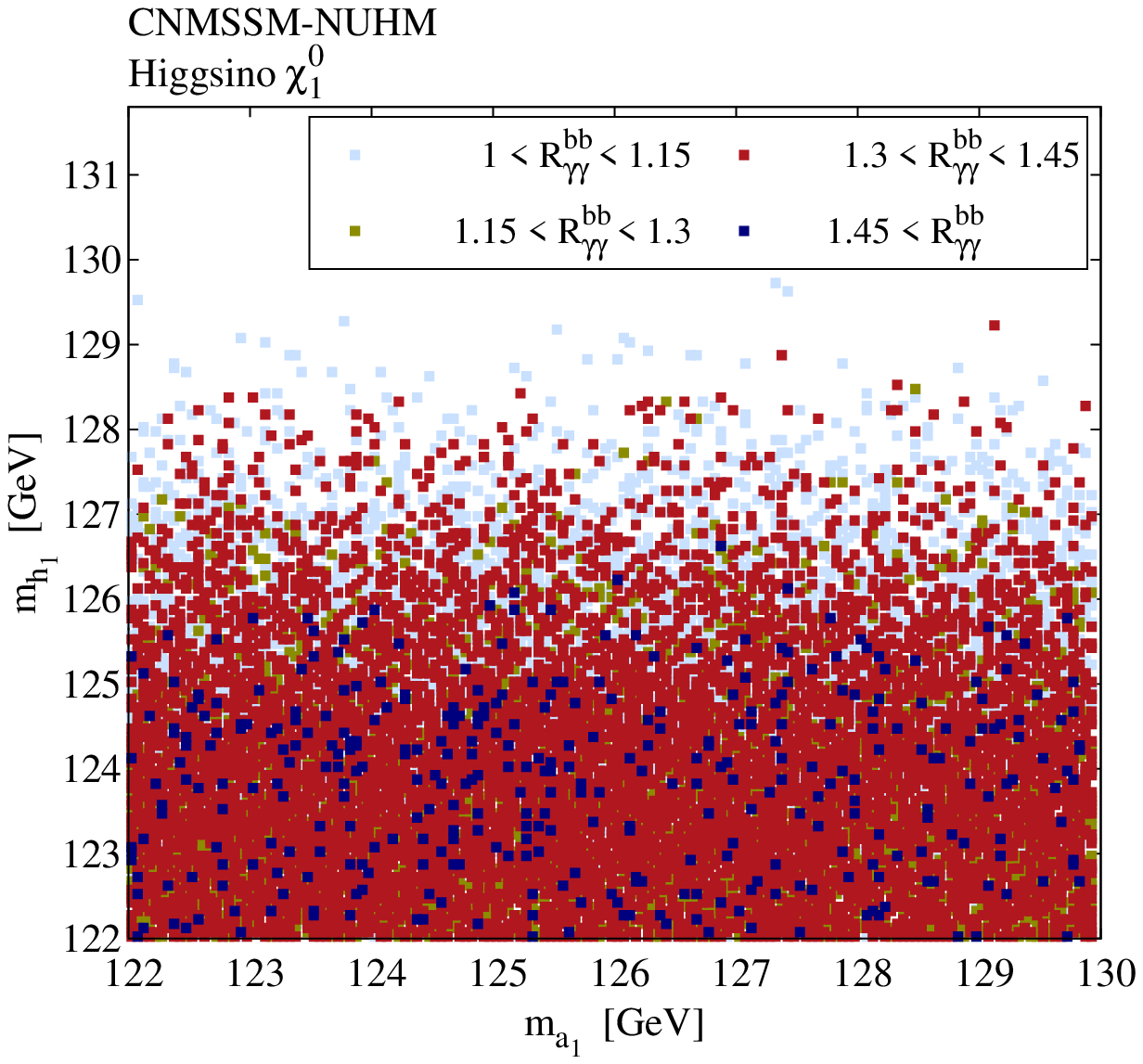}
}%
\caption[]{(a)-(d) Ranges of CNMSSM-NUHM paramaters corresponding to
  the Higgsino region. (e) $\xi\sigsip$ obtained for this region as a function
  of $m_\chi$, where $\xi = \Omega_\chi h^2 /\Omega_{\rm total}
  h^2$. (f) Ranges of $m_{h_1}$ and $m_{a_1}$ obtained in this
  region. See text for details.}
\label{fig:HIG}
\end{figure} 

A nearly pure Higgsino-like neutralino can generate large enough $\Omega_\chi h^2$
only if $m_{\chi} \simeq \mueff  \sim 1$\tev, but
such high values of \mueff\ will not yield the desired enhancement
in the $a_1 \rightarrow \gamma\gamma$ rate. Therefore, in order to
obtain a sizeable enhancement one has to relax the condition on neutralino
relic density (thereby allowing low \mueff\ and, therefore, $\Omega_\chi
h^2$ to be too low). One can assume that a neutralino contributes only
partially to the relic abundance of the universe beside some other DM
candidate, e.g., the axion. In that case $\Omega_\chi h^2= \xi\Omega _{\rm total} h
^2$, where $\xi$ is the fraction of the total relic
abundance produced by $\chi$ and $\Omega _{\rm total} h
^2 = 0.112$. 
Another possibility is that the entire relic abundance is
due to an alternative DM candidate particle in the model.
Often considered examples of such an additional/alternative DM candidate are
gravitino (see, e.g.,\cite{Bailly:2009pe,*Roszkowski:2012nq} for recent analyses in the MSSM) and/or
axino\cite{Covi:1999ty,*Covi:2001nw,*Choi:2007rh}. 
The first (second) of these candidates is (not) tightly constrained by Big Bang Nucleosynthesis 
but both are likely to be allowed in this region due to the low neutralino yield at
freeze-out. 

In Fig.\,\ref{fig:HIG}a we show the region in the (\mzero,\,\mhalf) plane
generating a light $a_1$ and a Higgsino-like $\chi$. 
The color assignment of the points is the same
 as in the singlino-Higgsino region.
Large values of \mzero\ are preferred again
in order to enhance $m_{h_1}$ through radiative corrections from the
SUSY sector. \mhalf\ also takes large values in order to minimize the
bino component of $\chi$ with large \mueff--\mhalf\ splitting. 
In Fig.\,\ref{fig:HIG}b we see that once again $\tan\beta$
spans a fairly wide range but small to intermediate values are
favored by $R^{bb}_{\gamma\gamma} (\rm a_1)$, for the same reasons as
in the singlino-Higgsino region. 
Larger values of negative \azero\ are favored so that $m_{h_1}$ can be
maximized.

In Fig.\,\ref{fig:HIG}c we show \akap\ vs \azero.  As in the
singlino-Higgsino region, for the (comparatively smaller) positive
\akap\ at the GUT scale only large negative \azero\ values are
allowed. However, in contrast with that region, a considerable number
of points is visible for negative \akap\ and negative \azero\ up to
$\sim$$-$2\tev. In this portion of the region negative \azero\ makes
\alam\ run upwards between the GUT scale and \msusy\ and, since the
latter is negative, it also drives \akap\ upwards to smaller negative
values.  Naturally then, negative \azero\ should not be very large or
\akap\ at \msusy\ will be driven positive.  Positive \azero\ solutions
are also possible, as long as they don't yield
positive \akap\ at \msusy.  In Fig.\,\ref{fig:HIG}d the ranges of the
paramaters \lam\ and \kap\ favored by our scenario under consideration
are shown for this region. A larger range of $\kappa$ is favored in
this region compared to the singlino-Higgsino region since $s$ is more
free to vary owing to the comparatively less constrained \akap\ (which
does not need to be as large to give correct $m_{a_1}$) $\lambda$ can
now take much smaller values than those allowed in the
singlino-Higgsino region but cannot be as large as there. This is in
fact the main feature distinguishing this region from the
singlino-Higgsino region in terms of the parameter space of the
model. Evidently, larger enhancement in the $\gamma\gamma$ rate is
favoured by large values of $\lambda$.

In Fig.\,\ref{fig:HIG}e
we show the distribution of the points in the ($m_{\chi}$,\,$\xi\sigsip$)
plane for this region.\footnote{The figure assumes that $\chi$
 is only responsible for a small portion of the observed relic
 abundance. For the points obtained in the scan $\xi\leq 0.05$.} 
Almost all the points obtained in this region
lie below the XENON100 line and a portion of these points even lies
below the projected XENON1T line. Since $\chi$ is almost purely Higgsino here,
the smaller its mass the bigger the enhancement in $R^{bb}_{\gamma\gamma}
(a_1)$ is generated. 
Moreover, this region
corresponds to large values of \mzero\ and \mhalf\, so that the
squarks and gluinos are always much heavier than the current
LHC reach. Nevertheless, as discussed earlier, a more precise
measurement of $R^{bb}_{\gamma\gamma}$ could still introduce limits on $m_{\chi}$
and $m_{\chi_1^\pm}$. Such derived upper limits are, therefore, especially
interesting from the experimental point of view. 

Finally, again as a result of large
allowed values of negative \azero, $h_1$ as heavy as 129\gev\ can be obtained
in this region, as can be seen in Fig.\,\ref{fig:HIG}f. $a_1$ mass is
evenly distributed in the defined range, almost always showing a large
enhancement in the $\gamma\gamma$ rate.   
Finally, a majority of
points in this region show a big enhancement, up to $\sim$50\% above the SM
expectation, in the $\gamma\gamma$ rate.
We also note here that both \brbsmumu\ and \brbxsgamma\ always lie around their respective SM values.

\subsubsection{Focus-point region}

\begin{figure}[p]
\centering
\subfloat[]{%
\label{fig:-a}%
\includegraphics*[height=6.5cm]{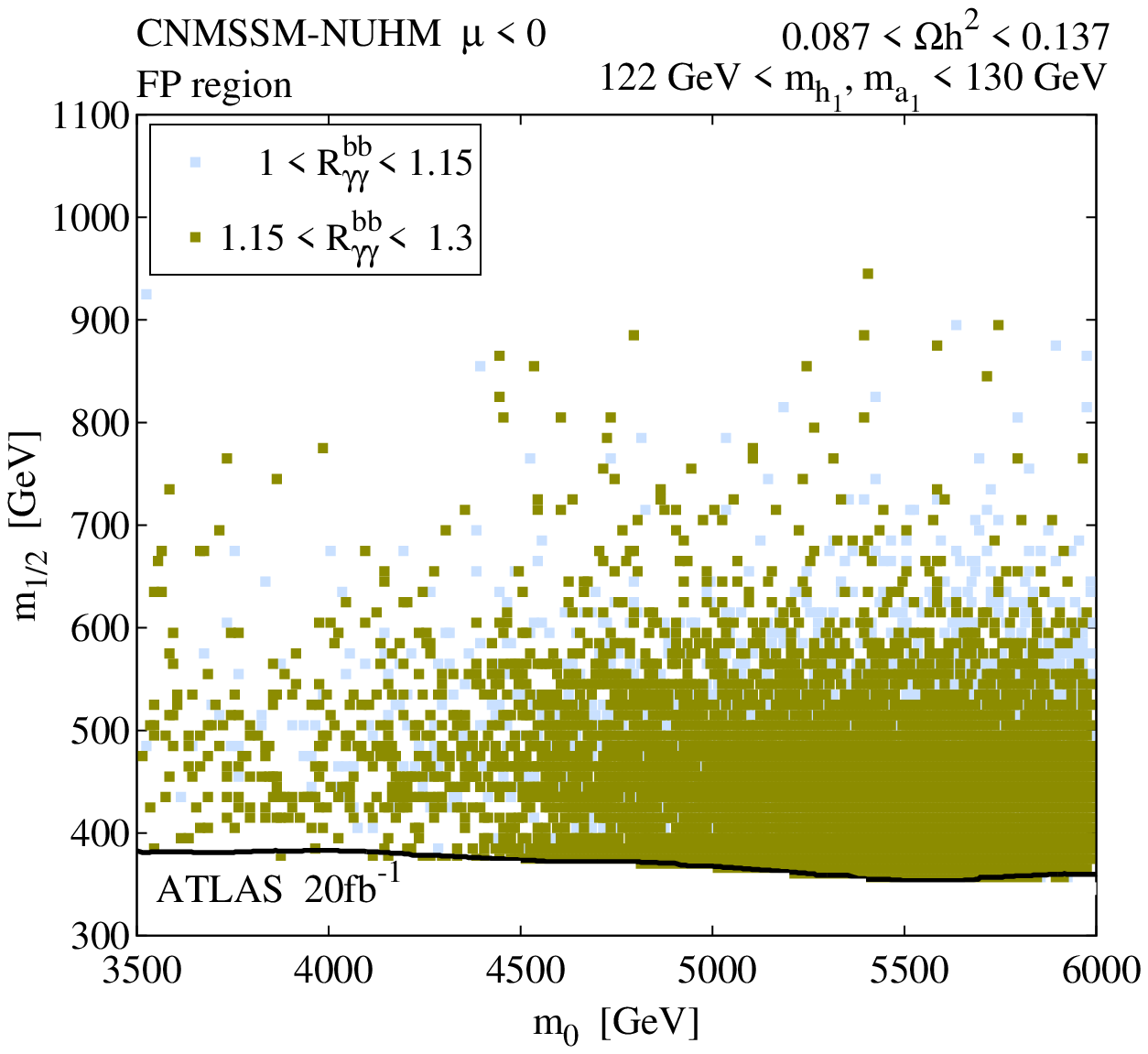}
}%
\hspace{0.5cm}%
\subfloat[]{%
\label{fig:-b}%
\includegraphics*[height=6.5cm]{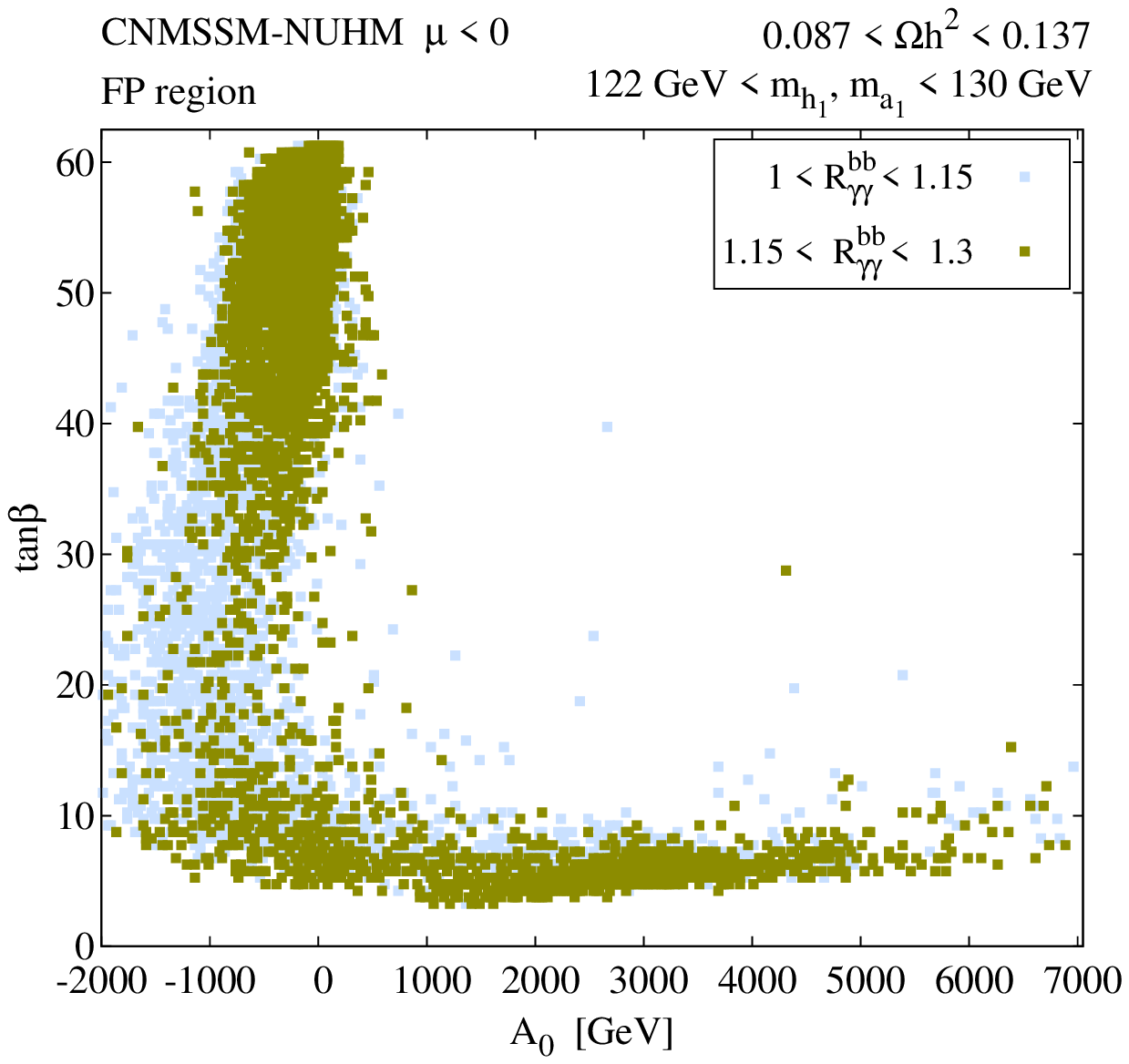}
}%

\subfloat[]{%
\label{fig:-a}%
\includegraphics*[height=6.5cm]{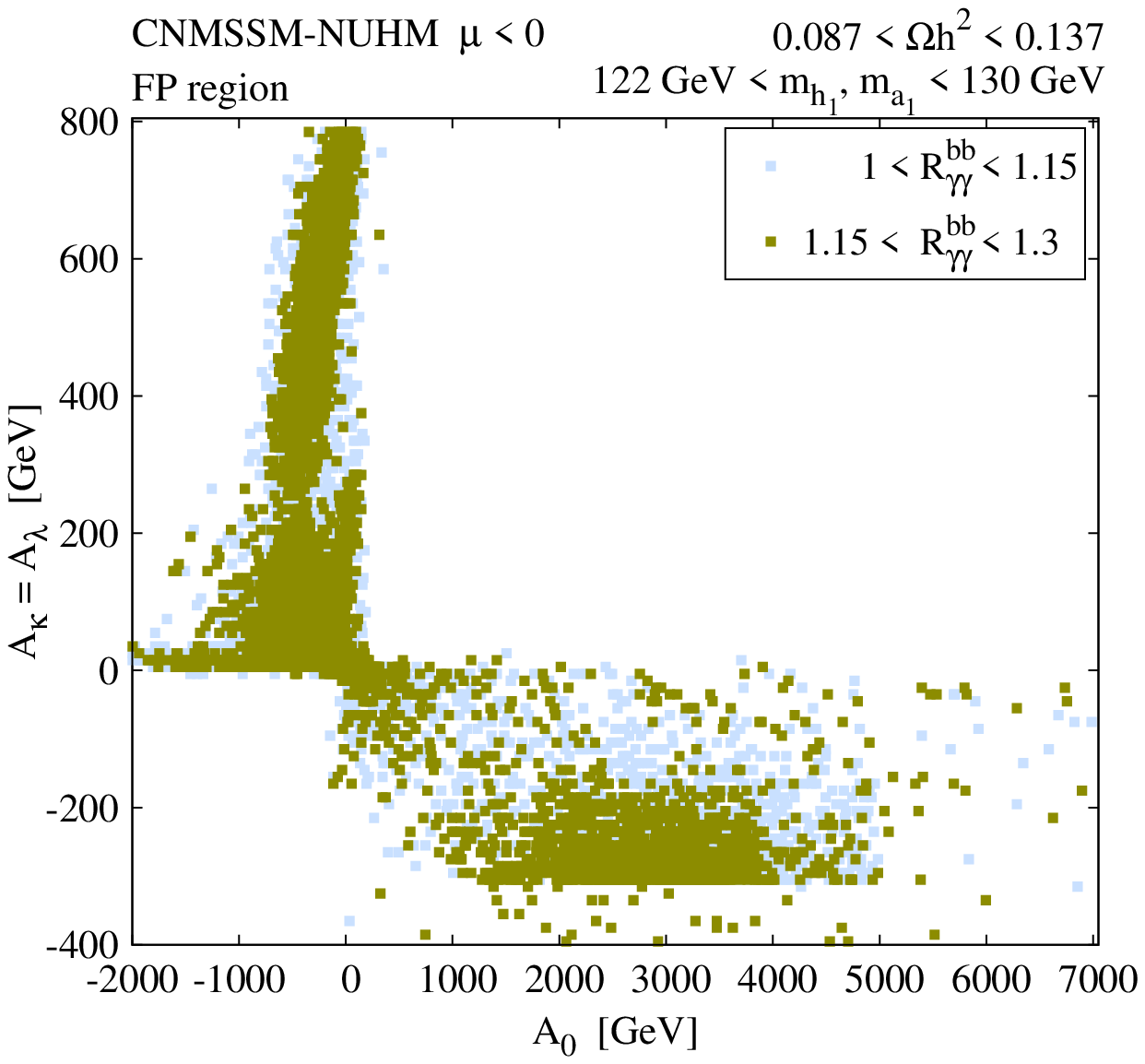}
}%
\hspace{0.5cm}%
\subfloat[]{%
\label{fig:-b}%
\includegraphics*[height=6.5cm]{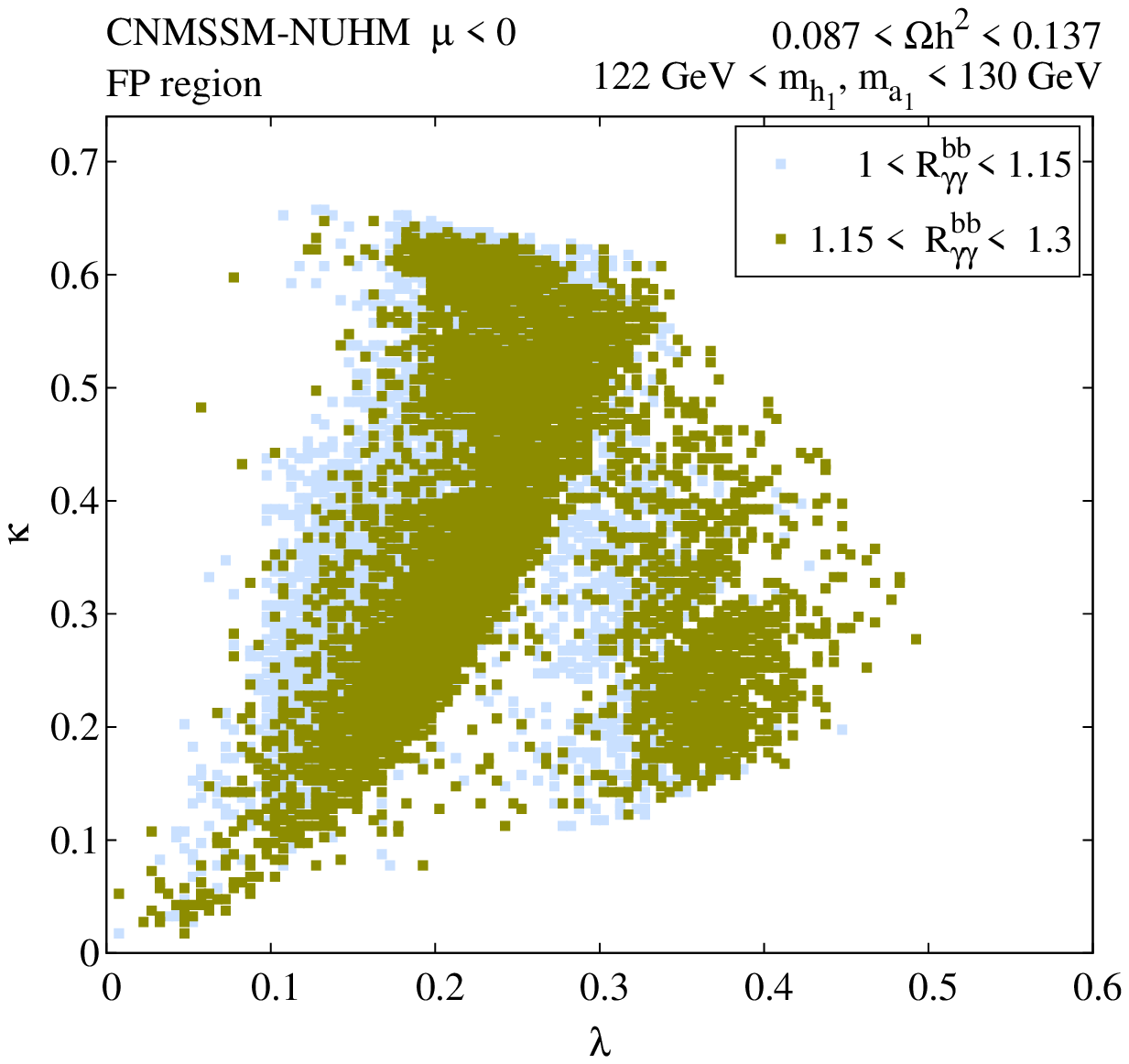}
}%

\subfloat[]{%
\label{fig:-a}%
\includegraphics*[height=6.5cm]{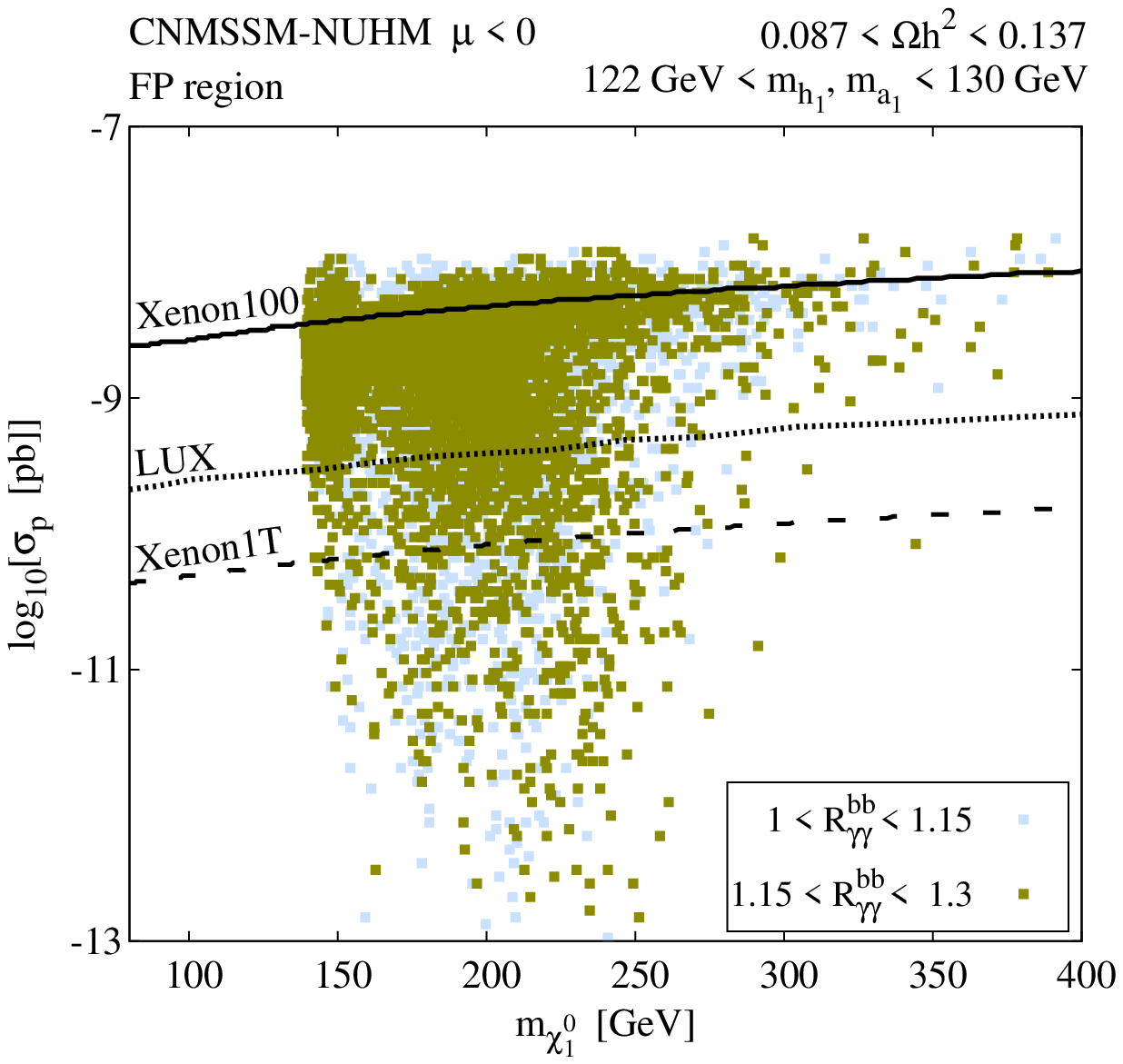}
}%
\hspace{0.5cm}%
\subfloat[]{%
\label{fig:-b}%
\includegraphics*[height=6.5cm]{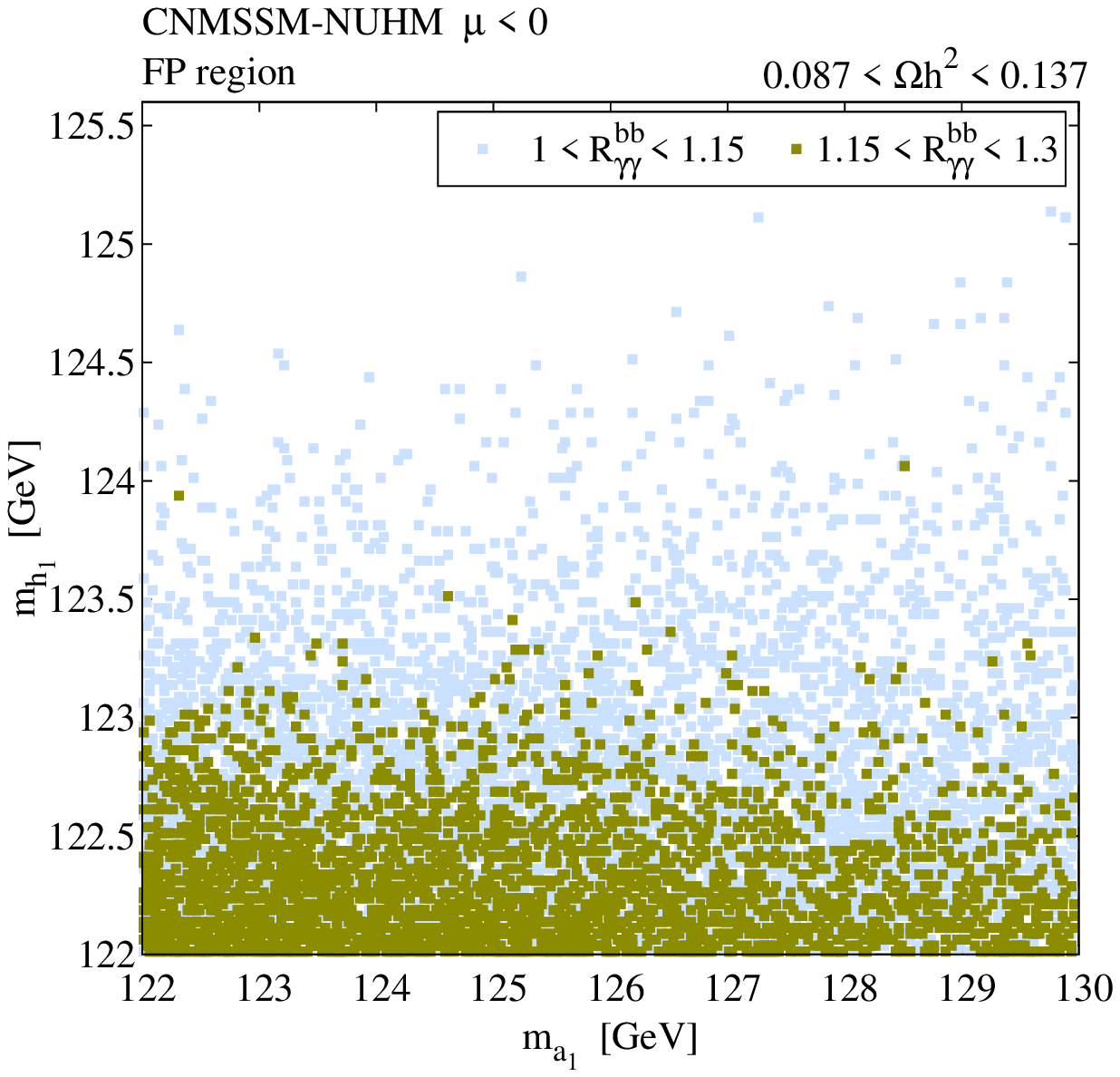}
}
\caption[]{(a)-(d) Ranges of CNMSSM-NUHM paramaters corresponding to
  the FP region. (e) \sigsip\ obtained for this region as a function
  of $m_\chi$. (f) Ranges of $m_{h_1}$ and $m_{a_1}$ obtained in this
  region. See text for details.}
\label{fig:FP}
\end{figure}

A light neutralino with mixed bino-Higgsino composition can generate
correct DM relic density, $\Omega_\chi h^2$, in the so called focus-point (FP)
region of minimal SUSY models in
general\cite{Chan:1997bi,*Feng:1999zg,*Feng:2000gh}. 
Since this region satisfies
the constraints from XENON100 and
\brbxsgamma\ measurement better  when $\mueff < 0$\cite{Roszkowski:2007fd}, we
shall pursue this case here. 
In Fig.\,\ref{fig:FP}a we show the region in the (\mzero,\,\mhalf) plane
generating a light $a_1$ ($122\gev \leq m_{a_1} \leq 130\gev$) and $\chi$ with a dominant 
bino and a small Higgsino component.
We see that while large values of \mzero\ are
favored in order to enhance the mass of $h_1$, \mhalf\ is typically
low, which is necessary for producing a mixed bino-Higgsino
$\chi$. This region, however, lies very close to the current 95\%
CL exclusion limit from ATLAS and should potentially be
tested soon. 

In Fig.\,\ref{fig:FP}b we show the favored ranges of the \azero\ and
$\tan\beta$ parameters.  $\tan\beta$ is almost always $\gtrsim 5$ to
allow enhancement in the $h_1b\bar{b}$ coupling in our considered
Higgs production mode, as noted earlier. However, we see in the figure
that for high positive \azero\ $\tan\beta$ is limited to small values,
$\lesssim 15$. The reason is that large \tanb\ results in an enhanced
Yukawa coupling of $h_1$ to $b\bar{b}$ and
$\tau\bar{\tau}$. Consequently \alam\ runs downwards rapidly from its
GUT scale value (we shall see below that large positive \azero\
coincides with negative \alam) to more negative values at \msusy.
This in turn causes \akap\ to run upwards from its GUT scale value,
raising $m_{a_1}$ beyond the desired range. The effective upper bound
on \tanb\ is relaxed for lower $|A_0|$, when the running is slower. We
point out here that the opposite effect of $\tan\beta$ was noted in
the other two regions due to the fact that there \azero\ was negative
which made \alam\ and \akap\ run in the opposite directions to those
here.

In Fig.\,\ref{fig:FP}c we show the distribution of the parameters
\azero\ and $A_\kappa = A_\lambda$ at the GUT scale. 
We notice that \azero\ stops at much smaller negative values than it
would be expected to take in order to maximize $m_{h_1}$. 
This is due to the reason explained in \refsec{Pseudo}. The main contribution to $m_{a_1}$
comes from the leading term in Eq.\,(\ref{eq:ma1}). Since in this
region we assume $\mueff<0$, \azero\ is strongly bounded from below in
order to minimize the effect of the second term there. Such negative \azero,
by causing positive \alam\ to run upwards, also pushes \akap\
to somewhat large positive values at the GUT scale resulting in its small positive
values at \msusy, since it runs in the opposite direction to
\alam. This is also the reason why no points are visible in the region
with negative \akap\ and negative \azero as opposed to the Higgsino
region, but conversely one can see some points with positive \akap\ and positive \azero.
Finally, for negative \akap\ large positive \azero\ can be reached, since such values
of \azero\ drive positive \alam\ at the
GUT scale downwards, which in turn causes \akap\ to run upwards to
positive values at \msusy. 

Fig.\,\ref{fig:FP}d shows the ranges of \lam\ and \kap\ corresponding
this scenario. The distribution of these two parameters almost mimics
that in the Higgsino region and as in there larger values of
enhancement in $R_{\gamma\gamma}^{bb}
(h_1+a_1)$ are obtained for large $\lambda$.
In Fig.\,\ref{fig:FP}e we show how the FP region fares against the XENON100 limits. 
Note that $m_{\chi}$ is bounded
from below by the ATLAS limit on \mhalf\ in this region as it is
bino-dominated.
We see that a majority of the allowed points with
an enhanced $\gamma\gamma$ rate lie below the XENON100 line. Most of
this region, however, lies above the LUX limit, while the XENON1T
data should be able to test almost all of it.  
Fig.\,\ref{fig:FP}f shows the allowed masses of $h_1$ and $a_1$
in this region. We see that $m_{h_1}$ is always lighter than 124\gev,
which is a consequence of $\mueff<0$  
not allowing very large values of negative \azero\ in this
region, as discussed above. $a_1$, on the other hand, can easily have a mass
greater than 125\gev.

Overall, we notice only a relatively small enhancement, up to
$\sim$25\% in the $\gamma\gamma$ rate compared to the SM expectation
in this region of the CNMSSM-NUHM parameter space. The reason for this
is that $m_{\chi_1^\pm}$ is not allowed to take values even as small
as those possible in the Higgsino region due to the lower bound on the
mass of $\chi$ discussed above ($m_{\chi_1^\pm} \simeq \mu_{\rm eff} >
m_{\chi}$).  \brbsmumu\ in this region varies between $2\times
10^{-9}$ and $5.5\times 10^{-9}$, which is within $2\sigma$ of the
experimentally measured value $3.2 \times 10^{-9}$, taking into
account the theoretical error (as in\cite{Kowalska:2012gs}).  On the
other hand, \brbxsgamma\ takes values between $3.1\times 10^{-4}$ and
$3.7\times 10^{-4}$ and hence is always close to the experimental
value. This region, owing mainly to the facts that $m_{h_1}$ finds it
difficult to reach the experimentally observed value and that the
reduced $\gamma\gamma$ rate of $a_1$ barely exceeds 0.25, is
the least favored of the three regions explored here. \\

\begin{figure}[h!]
\centering
\subfloat[]{%
\label{fig:-a}%
\includegraphics*[height=7cm]{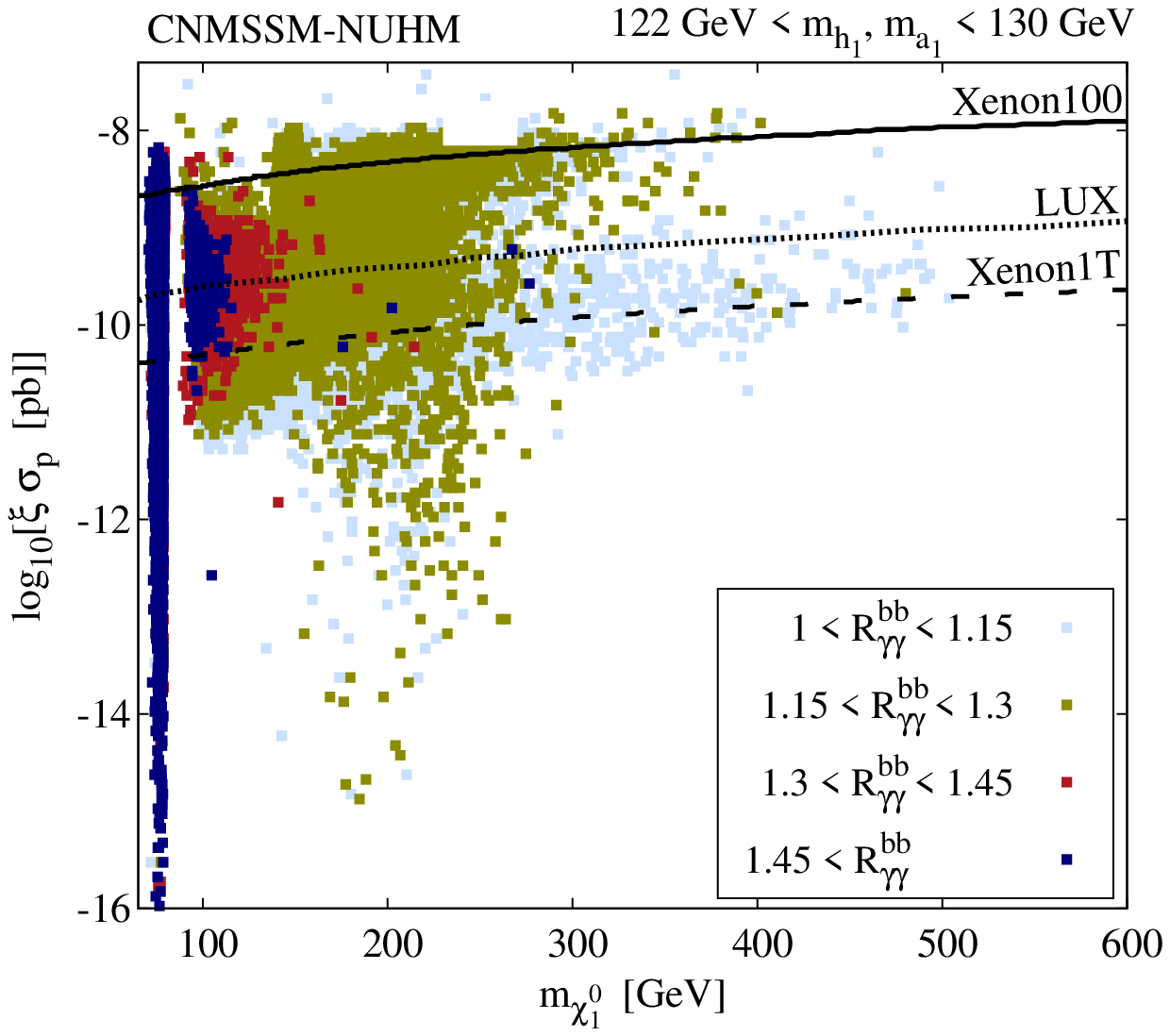}
}%
\hspace{0.5cm}%
\subfloat[]{%
\label{fig:-b}%
\includegraphics*[height=7cm]{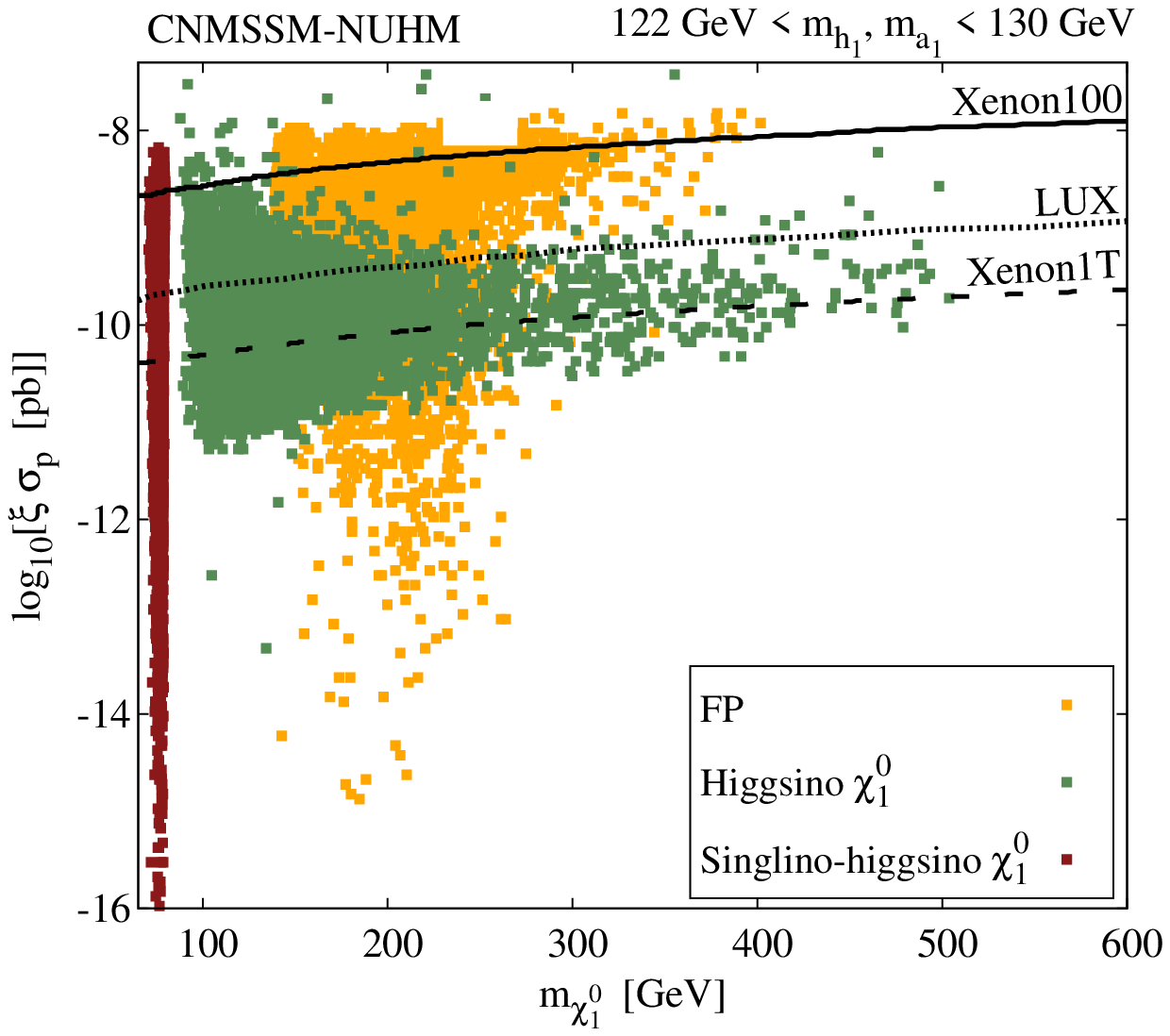}
}
\caption[]{ (a) The range of \sigsip\ for giving an enhancement in $R_{\gamma\gamma}^{bb}
(h_1+a_1)$ versus the neutralino mass $m_\chi$. Also shown are the 90\%
CL exclusion limits from XENON100 as
well as the 90\% CL limits expected from
the LUX and XENON1T experiments. $\xi = \Omega_\chi h^2
/\Omega_{\rm total}  h^2$ when $\chi$ is almost purely Higgsino but 1 otherwise.
(b) The  ($m_\chi$,\,\sigsip)
plane showing the three CNMSSM-NUHM regions where $R_{\gamma\gamma}^{bb}
(h_1+a_1)$ is enhanced. Maroon squares denote the singlino-Higgsino region, green squares the
Higgsino region and yellow squares the FP region. $\xi = \Omega_\chi h^2
/\Omega_{\rm total}  h^2$ in the Higgsino region, but 1 in the
singlino-Higgsino and FP regions.} 
\label{fig:combined}
\end{figure}

To summarize, in Fig.\,\ref{fig:combined}a we show the
range of $m_\chi$ across all the regions for which an enhancement in $R_{\gamma\gamma}^{bb}
(h_1+a_1)$ was obtained in our CNMSSM-NUHM scan, and its compatibility 
with the current and expected limits on \sigsip. 
These regions are identified seperately in
Fig.\,\ref{fig:combined}b, again, in the  ($m_\chi$,\,$\xi\sigsip$)
plane, where the orange squares denote the FP region, dark green squares the
Higgsino region and brown squares the singlino-Higgsino region. 

\begin{figure}
\centering
\subfloat[]{%
\label{fig:-a}%
\includegraphics*[height=7cm]{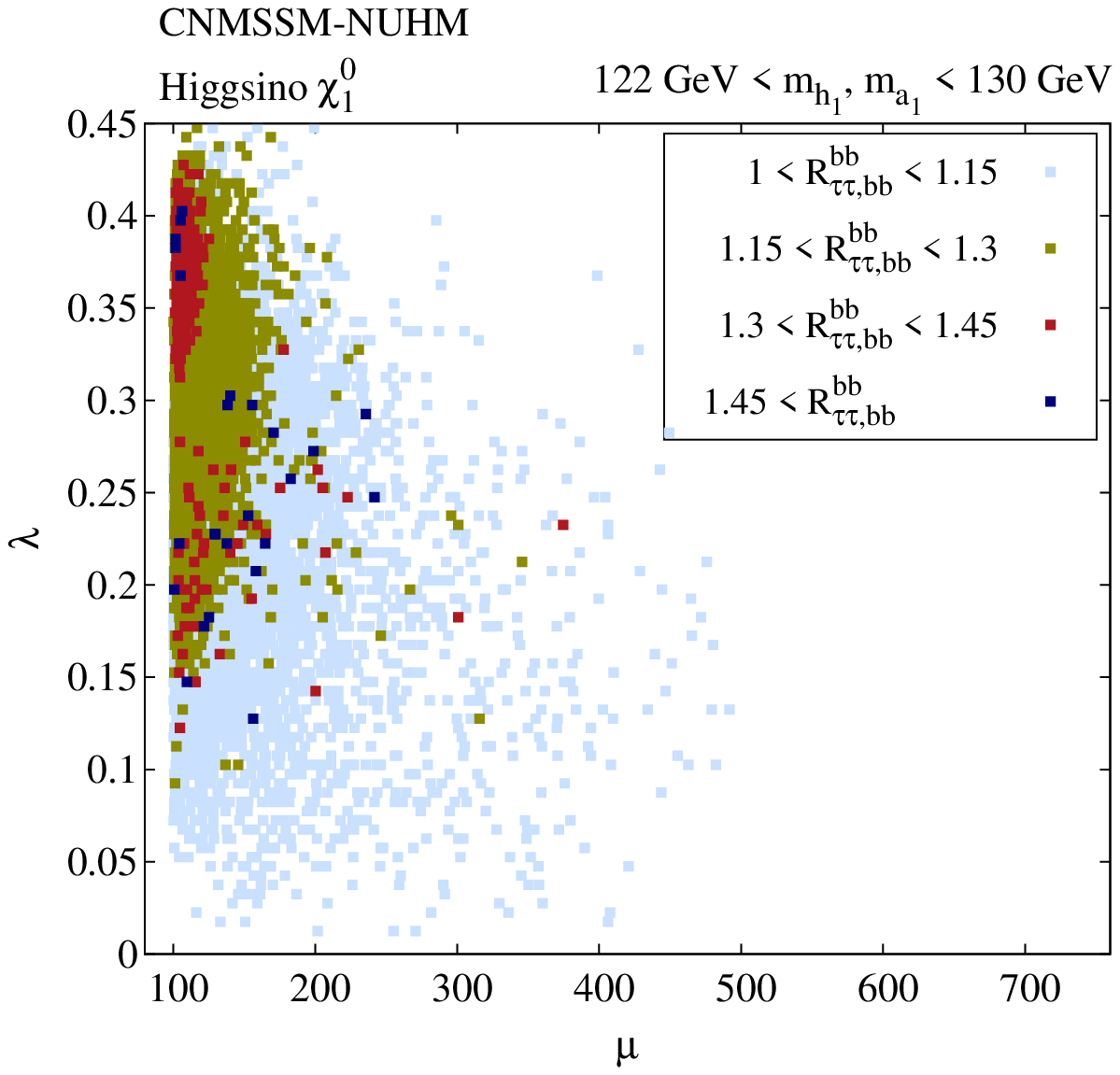}
}%
\hspace{0.5cm}%
\subfloat[]{%
\label{fig:-b}%
\includegraphics*[height=7cm]{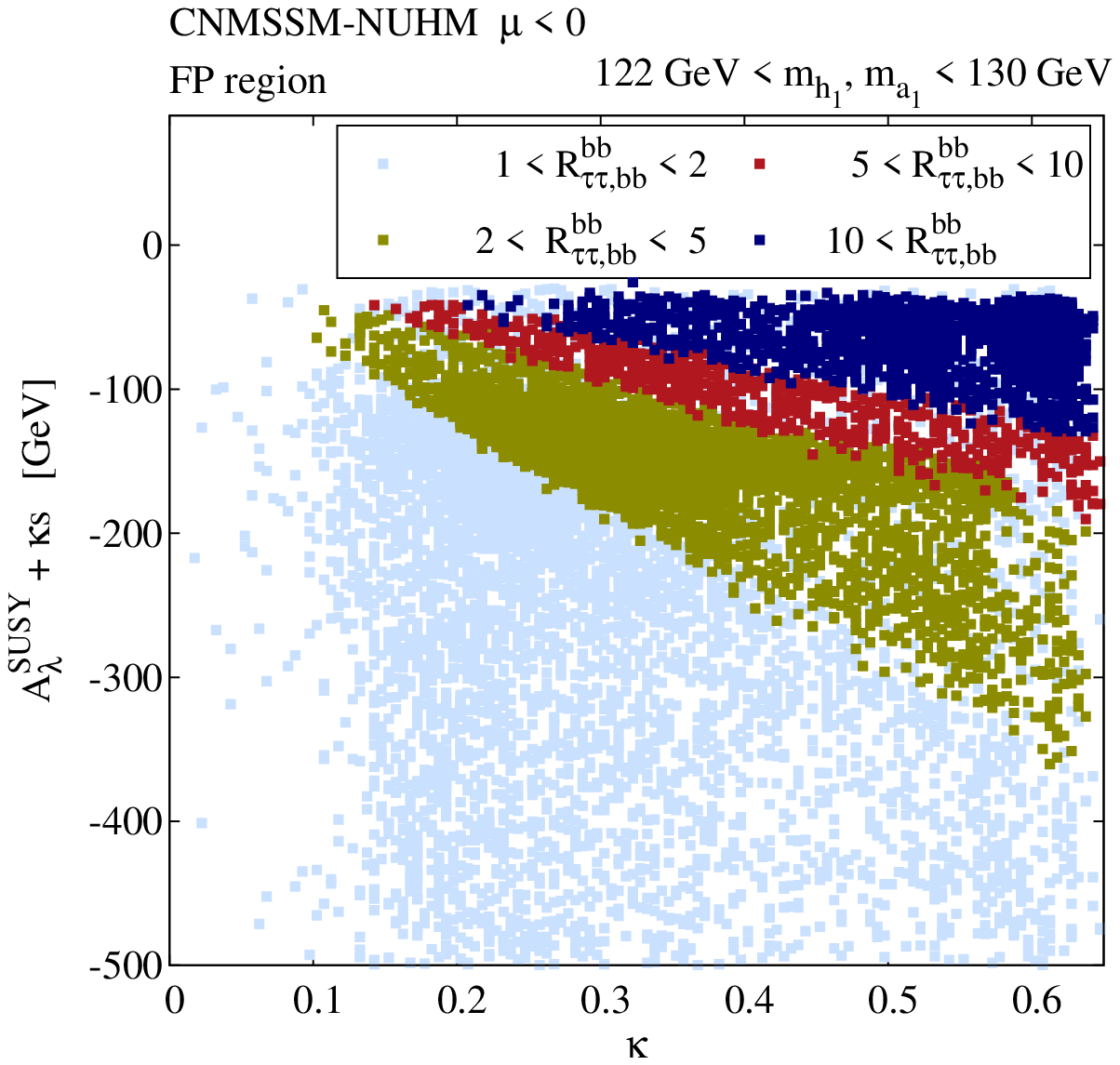}
}%
\caption[]{(a) Enhancement in $R_{b\bar{b}/\tau^+\tau^-}^{bb}
(h_1+a_1)$ obtained in the Higgsino region as a function of the \lam\ and
\mueff\ parameters. (b) Enhancement in $R_{b\bar{b}/\tau^+\tau^-}^{bb}
(h_1+a_1)$ obtained in the FP region as a function of \kap\ and
$(A^{\rm SUSY}_{\lambda}+\kappa s)$ from the denominator of $|P_{11}''|$. See text for details.}
\label{fig:rtautau}
\end{figure} 

 \subsection{$b\bar{b}/\tau^+\tau^-$ rate enhancement}

In this subsection we highlight only the important features corresponding to
the $b\bar{b}$ and $\tau^+\tau^-$ decay channels of $a_1$ for the three regions discussed in
detail above.  As noted in \refsec{sec:a1tautau}, contrary to the case of a
MSSM-like scalar Higgs boson,
$\tan\beta$ affects the $b\bar{b}/\tau^+\tau^-$ rate of $a_1$ only indirectly through the term
in the denominator of Eq.\,(\ref{eq:rcs_tau}). 
Instead, a sizable $R_{b\bar{b}/\tau^+\tau^-}^{bb}
(a_1)$ is an additional consequence of the conditions necessary
to obtain an enhancement in $R_{\gamma\gamma}^{bb}(a_1)$, i.e., large
\lam\ and small \mueff. This is demonstrated in
Fig.\,\ref{fig:rtautau}(a) for the Higgsino region, where one can see that
the enhancement in $R_{b\bar{b}/\tau^+\tau^-}^{bb}
(h_1+a_1)$ rises with increasing \lam\ and decreasing \mueff. 
In the singlino-Higgsino region (not shown in
the figure) $R_{b\bar{b}/\tau^+\tau^-}^{bb}
(h_1+a_1)$ is always larger than 1.6  for the entire range of \lam, seen in
Fig.\,\ref{fig:SIN}(d), and can be as high as 1.9. It will therefore
result in a small blue region at the top left corner of Fig.\,\ref{fig:rtautau}(a). 

In the FP region $R_{b\bar{b}/\tau^+\tau^-}^{bb}
(a_1)$ can in fact have extremely large values, $\sim$100. However, this should not be interpreted as
a characteristic feature specific to the FP region, but as a result of negative \mueff\ assumed for this
region. $R_{b\bar{b}/\tau^+\tau^-}^{bb}
(a_1)$ increases as the denominator, $A_\lam^{\rm
  SUSY} + \kappa s$, of $|P_{11}''|$ in Eq.\,(\ref{eq:rcs_tau})
approaches zero. For small negative \mueff\ and large positive \lam, resulting in small
negative $s$, the size of the denominator reduces as $\kappa$ grows. In
Fig.\,\ref{fig:rtautau}(b) we show how $R_{b\bar{b}/\tau^+\tau^-}^{bb}
(h_1+a_1)$ enhances with increasing $\kappa$ and decreasing value of
the above denominator term, and can acquire a huge value before the
perturbative upper limit on the former is reached. 
Evidently a similar effect of negative \mueff\ should manifest in the
other two regions also. However, since negative \mueff\ causes a
tension between $m_{a_1}$ and $m_{h_1}$ and does not allow both of
these to be around 125\gev, as discussed in detail in \refsec{sec:a1mass} and
as noted in the FP region, we retain $\mueff>0$ in the
singlino-Higgsino and the Higgsino regions. We thus expect the
enhancement in the $b\bar{b}/\tau^+\tau^-$ channels to be larger in these two regions
also for $\mueff<0$, but at the cost of $m_{h_1}$ and \amu\ lying
far from their respective experimentally measured values. 

\section{\label{Summary}Summary}

We have proposed an experimental test of a scenario in the NMSSM in
which the lightest pseudoscalar of the model as well a SM-like
lightest scalar boson both have masses around $\sim$125\gev. The
pseudoscalar could be distinguishable from the scalar at the LHC in
the associated Higgs production mode with a $b\bar{b}$ pair in the
final state. This is because it will contribute significantly to the
observed signal rate in the $\gamma\gamma$ and $b\bar{b}/\tau^+\tau^-$
channels but, since a pseudoscalar doesn't couple to $W$ and $Z$
bosons, the measured rate in the $WW/ZZ$ channels will be due only to
the scalar and therefore SM-like. We have discussed the conditions
necessary to obtain $a_1$ with the correct mass and noted that an
observable enhancement in its $\gamma\gamma$ decay rate is made
possible by a light chargino entering in its one-loop effective
coupling to two photons. We have also discussed in detail how the
conditions to obtain such a light chargino in turn lead to an
enhancement in the $b\bar{b}$ and $\tau^+\tau^-$ rates also.  We have
argued that, due to very specific requirements on the composition of
$a_1$, which should be singlet-like, and of the light chargino, which
should be almost purely Higgsino-like, such a scenario cannot be
realized in the MSSM and is extremely unlikely in the fully
constrained NMSSM.

We have, therefore, analyzed the CNMSSM with the universality
conditions lifted in the Higgs sector to study the 
scenario at hand. We have scanned the parameter space of this model in order
to look for regions that can allow both $\chi_1^\pm$ and $a_1$
with the desired masses and compositions. We have found that these regions can be divided into
three broad types based on the composition of the
neutralino which, owing to the condition on $\chi_1^\pm$, should also have
a large Higgsino component. These regions include the
singlino-Higgsino region, where $\chi$ is Higgsino-dominated but 
with an admixture of the singlino, the Higgsino region, where it is almost purely
Higgsino, and the FP region, where it is a
bino-Higgsino mixture. The region showing least enhancement in the
$\gamma\gamma$ rate of $a_1$ is the FP region, where it only reaches
up to $\sim$25\%, while the most favored one is
the mixed Higgsino-singlino region, where the enhancement can be as high as
$\sim$60\%. However, we noted that for negative \mueff\ the FP region
satisfies the constraints from \sigsip\ and  \brbxsgamma\ better and
can also have very large enhancement in the $b\bar{b}$ and $\tau^+\tau^-$
rates, but cannot yield $m_{h_1}$ greater than 124\gev.
   
We have also stressed the fact that such a singlet-like $a_1$ is likely to remain invisible at
the LHC in the gluon-fusion production channel. The reason is that
while the effective coupling of $a_1$ to the $\gamma\gamma$, 
 $b\bar{b}$ and $\tau^+\tau^-$ pairs gets
enhanced, the effective coupling to two gluons is still highly
suppressed compared to a SM-like Higgs boson. We have, therefore,
emphasized that a more focussed analysis of the associated Higgs boson
production mode with $b\bar{b}$ pair, which is the least favorable production mode
for a SM-like Higgs boson, is essential. By revealing such a pseudoscalar
through the `triple enhancement' in its decay rates,
this production mode could provide 
a clear signature of our considered NMSSM scenario, in
particular, and of beyond the SM physics, in general. 


\begin{center}
  \textbf{ACKNOWLEDGMENTS}
\end{center}

  \noindent The authors would like to thank Y.-L.~Sming Tsai for valuable
  discussions and inputs. They are also thankful to A.~Kalinowski and
  M.~Pierini for their important remarks about the $b\bar{b}h$
  production mode. This work has been funded in part by the Welcome Programme
  of the Foundation for Polish Science. 
  L.R. is also supported in part by the Polish National Science Centre
  grant N N202 167440, an STFC consortium grant of Lancaster,
  Manchester and Sheffield Universities and by the EC 6th Framework
  Programme MRTN-CT-2006-035505. The use of the CIS computer cluster at NCBJ is gratefully acknowledged. 

\bibliographystyle{utphysmcite}	

\bibliography{myref}

\end{document}

%% file: macros_bf4.tex
\newcommand{\newc}{\newcommand*}

\long\def\begincomment#1\endcomment{%
        \begingroup\sf\baselineskip12pt#1\endgroup}

\newc{\etal}{\textrm{et al.}} 
\newc{\eg}{\textrm{e.g.}} 
\newc{\ie}{\textrm{i.e.}}
\newc{\etc}{\textrm{etc}}
\newc\vs{\textrm{vs.}}
\newc{\cl}{\rm {CL}}

\newc{\ev}{\ensuremath{\,\mathrm{eV}}}
\newc{\kev}{\ensuremath{\,\mathrm{keV}}}
\newc{\mev}{\ensuremath{\,\mathrm{MeV}}}
\newc{\gev}{\ensuremath{\,\mathrm{GeV}}}
\newc{\tev}{\ensuremath{\,\mathrm{TeV}}}

\newc{\MeV}{\mev} 
\newc{\TeV}{\tev}
\newc{\invpb}{\ensuremath{/\text{pb}}}
\newc{\invfb}{\ensuremath{/\text{fb}}}

\newc\nb{\ensuremath{\,\mathrm{nb}}} \newc\pb{\ensuremath{\,\mathrm{pb}}} \newc\fb{\ensuremath{\,\mathrm{fb}}}

\newc\pc{\ensuremath{\,\mathrm{pc}}}
\newc\kpc{\ensuremath{\,\mathrm{kpc}}}
\newc\mpc{\ensuremath{\,\mathrm{Mpc}}}

\newc\ps{\ensuremath{\,\mathrm{ps}}} 

\newc\cmeter{\ensuremath{\,\mathrm{cm}}} 
\newc\meter{\ensuremath{\,\mathrm{m}}} 
\newc\kmeter{\ensuremath{\,\mathrm{km}}}

\newc\second{\ensuremath{\,\mathrm{s}}}
\newc\msecond{\ensuremath{\,\mathrm{ms}}}
\newc\nsecond{\ensuremath{\,\mathrm{ns}}}
\newc\psecond{\ensuremath{\,\mathrm{ps}}}

\newc{\chisqmin}{\ensuremath{\chi^2_{\mathrm{min}}}}
\newc{\Delchisq}{\ensuremath{\Delta\chi^2}}
\newc{\chisq}{\ensuremath{\chi^2}}

\newc{\like}{\ensuremath{\mathcal{L}}}

\newc\lsim{\ensuremath{\mathrel{\rlap{\lower4pt\hbox{\hskip1pt$\sim$}}\raise1pt\hbox{$<$}}}}
\newc\gsim{\ensuremath{\mathrel{\rlap{\lower4pt\hbox{\hskip1pt$\sim$}}\raise1pt\hbox{$>$}}}}

\newc{\VEV}[1]{\ensuremath{\langle #1 \rangle}}

\newc{\dl}{\ensuremath{\stackrel{\leftarrow}{D}}}
\newc{\dr}{\ensuremath{\stackrel{\rightarrow}{D}}}

\newc{\bcenter}{\begin{center}}    \newc{\ecenter}{\end{center}}
\newc{\bfl}{\begin{flushleft}}    \newc{\efl}{\end{flushleft}}
\newc{\bfr}{\begin{flushright}}    \newc{\efr}{\end{flushright}}

\newc{\bi}{\begin{itemize}}
\newc{\ei}{\end{itemize}}
\newc{\bed}{\begin{description}}
\newc{\eed}{\end{description}}
\newc{\ben}{\begin{enumerate}}
\newc{\een}{\end{enumerate}}

\newc{\be}{\begin{equation}}
\newc{\ee}{\end{equation}}
\newc{\bea}{\begin{eqnarray}}
\newc{\eea}{\end{eqnarray}}
\newc{\ra}{\rightarrow}

\newc{\alphas}{\ensuremath{\alpha_s}}
\newc{\alphatwo}{\ensuremath{\alpha_2}}
\newc{\alphaone}{\ensuremath{\alpha_1}}
\newc{\alphai}[1]{\ensuremath{\alpha_{#1}}}
\newc{\alphaem}{\ensuremath{\alpha_{\mathrm{em}}}}

\newc{\alphaeff}{\ensuremath{\alpha_{\mathrm{eff}}}}

\newc{\sineff}{\ensuremath{\sin \theta_{\mathrm{eff}}}}
\newc{\sinsqeff}{\ensuremath{\sin^2 \theta_{\mathrm{eff}}}}
\newc{\dalphahad}{\ensuremath{\Delta \alpha_{\mathrm{had}}}}

\newc{\yt}{\ensuremath{h_t}} \newc{\yb}{\ensuremath{h_b}} \newc{\ytau}{\ensuremath{h_{\tau}}}

\newc\mz{\ensuremath{M_Z}} 
\newc\mw{\ensuremath{m_W}}
\newc\mZ{\mz}        \newc\mW{\mw}

\newc\mhsm{\ensuremath{ m_{H_{\mathrm{SM}}}}}

\newc{\mtop}{\ensuremath{ m_t}}               \newc{\mtpole}{\ensuremath{ M_t}}
\newc{\mbottom}{\ensuremath{ m_b}} 
\newc{\mtau}{\ensuremath{ m_{\tau}}}
\newc{\mt}{\mtpole}
\newc{\mb}{\mbottom} 

\newc{\rtwogg}{\ensuremath{R_{h_2}(\gamma\gamma)}}
\newc{\rtwozz}{\ensuremath{R_{h_2}(ZZ)}}
\newc{\ronegg}{\ensuremath{R_{h_1}(\gamma\gamma)}}
\newc{\ronezz}{\ensuremath{R_{h_1}(ZZ)}}
\newc{\rsiggg}{\ensuremath{R_{h_\textrm{sig}}(\gamma\gamma)}}
\newc{\rsigzz}{\ensuremath{R_{h_\textrm{sig}}(ZZ)}}

\newc{\llbar}{\ensuremath{\ell\bar{\ell}}}
\newc{\tauptaum}{\ensuremath{ \tau^+\tau^-}}

\newc{\qqbar}{\ensuremath{ q\bar{q}}} \newc{\ppbar}{\ensuremath{ p\bar{p}}}
\newc{\bbbar}{\ensuremath{ b\bar{b}}} \newc{\ttbar}{\ensuremath{ t\bar{t}}}
\newc{\ffbar}{\ensuremath{ f\bar{f}}} \newc{\tautaubar}{\ensuremath{ \tau\bar{\tau}}}

\newc{\mchi}{\ensuremath{m_\neutone}}
\newc{\squark}{\ensuremath{\tilde{q}}}
\newc{\slepton}{\ensuremath{\tilde{l}}}
\newc{\gluino}{\ensuremath{\tilde{g}}} 
\newc{\mgluino}{\ensuremath{{m_{\gluino}}}}

\newc{\sthw}{\ensuremath{ \sin\theta_W}}              \newc{\cthw}{\ensuremath{\cos\theta_W}}
\newc{\tanthw}{\ensuremath{ \tan\theta_W}}              \newc{\cotthw}{\ensuremath{\cot\theta_W}}

\newc{\ssqthw}{\ensuremath{\sin^2 \theta_W}}

\newc{\msbar}{\ensuremath{\overline{MS}}} \newc{\drbar}{\ensuremath{\overline{DR}}}

\newc{\mtmtsmmsbar}{\ensuremath{ m_t(m_t)^{\msbar}_{{\mathrm{SM}}}}}
\newc{\mtmtsmdrbar}{\ensuremath{ m_t(m_t)^{\drbar}_{{\mathrm{SM}}}}}
\newc{\mtmtmssmdrbar}{\ensuremath{ m_t(m_t)^{\drbar}_{{\mathrm{SUSY}}}}}

\newc{\mbmbmsbar}{\ensuremath{ m_b(m_b)^{\msbar} }}

\newc{\mbmbsmmsbar}{\ensuremath{ m_b(m_b)^{\msbar}_{{\mathrm{SM}}}}}
\newc{\mbmzsmmsbar}{\ensuremath{ m_b(\mz)^{\msbar}_{{\mathrm{SM}}}}}
\newc{\mbmzsmdrbar}{\ensuremath{ m_b(\mz)^{\drbar}_{{\mathrm{SM}}}}}
\newc{\mbmzmssmdrbar}{\ensuremath{ m_b(\mz)^{\drbar}_{{\mathrm{SUSY}}}}}

\newc{\mtaumzsmmsbar}{\ensuremath{ m_{\tau}(\mz)^{\msbar}_{{\mathrm{SM}}}}}
\newc{\mtaumzsmdrbar}{\ensuremath{ m_{\tau}(\mz)^{\drbar}_{{\mathrm{SM}}}}}
\newc{\mtaumzmssmdrbar}{\ensuremath{ m_{\tau}(\mz)^{\drbar}_{{\mathrm{SUSY}}}}}

\newc{\alphasmzms}{\ensuremath{\alpha_s(M_Z)^{\overline{MS}}}}
\newc{\alphaimzms}[1]{\ensuremath{\alpha_{#1}(M_Z)^{\overline{MS}}}}

\newc{\alphaemmz}{\ensuremath{\alpha_{\mathrm{em}}(M_Z)^{\overline{MS}}}}

\newc{\mzero}{\ensuremath{{m_0}}}
\newc{\mhalf}{\ensuremath{ m_{1/2}}}
\newc{\tanb}{\ensuremath{\tan\beta}}
\newc{\azero}{\ensuremath{ A_0}}
\newc{\signmu}{\ensuremath{\rm{sgn}\,\mu}}
\newc{\atau}{\ensuremath{{A_{\tau}}}}

\newc{\mueff}{\ensuremath{\mu_{\rm{eff}}}}

\newc{\lam}{\ensuremath{{\lambda}}}
\newc{\kap}{\ensuremath{{\kappa}}}

\newc{\alam}{\ensuremath{{A_{\lambda}}}}
\newc{\akap}{\ensuremath{{A_{\kappa}}}}

 \newc{\hs}{\ensuremath{ H_s}}      
\newc{\mhs}{\ensuremath{ m_{H_s}}} 

\newc{\mgut}{\ensuremath{ M_{\rm GUT}}}
\newc{\mplanck}{\ensuremath{ M_{\rm P}}}      \newc{\mpl}{\ensuremath{ M_{\rm Pl}}}
\newc{\msusy}{\ensuremath{ M_{\rm SUSY}}}      \newc{\ms}{\ensuremath{ M_{\rm S}}}

 \newc{\hu}{\ensuremath{ H_u}}       \newc{\hd}{\ensuremath{ H_d}}
 \newc{\mhu}{\ensuremath{ m_{H_u}}}       \newc{\mhd}{\ensuremath{ m_{H_d}}}
 \newc{\mhuew}{\ensuremath{ m^{\ast}_{H_u}}}       \newc{\mhdew}{\ensuremath{ m^{\ast}_{H_d}}}
 \newc{\mhuewsq}{\ensuremath{ m^{\ast\, 2}_{H_u}}}       \newc{\mhdewsq}{\ensuremath{ m^{\ast\, 2}_{H_d}}}
 \newc{\mhl}{\ensuremath{m_\hl}} 
 \newc{\mhone}{\ensuremath{m_{h_1}}} 
 \newc{\mhtwo}{\ensuremath{m_{h_2}}} 
 \newc{\mglu}{\ensuremath{m_{\tilde g}}} 
 \newc{\mul}{\ensuremath{m_{\tilde{u}_L}}} 
 \newc{\mtone}{\ensuremath{m_{\tilde{t}_1}}} 
 \newc{\ma}{\ensuremath{m_A}} 
 \newc{\maone}{\ensuremath{m_{a_1}}} 
 \newc{\matwo}{\ensuremath{m_{a_2}}}
 \newc{\hone}{\ensuremath{h_1}}
 \newc{\htwo}{\ensuremath{h_2}}
 \newc{\aone}{\ensuremath{a_1}}
 \newc{\atwo}{\ensuremath{a_2}}

\newc{\sigsip}{\ensuremath{\sigma^{\rm SI}_{p}}}	\newc{\sigsin}{\ensuremath{\sigma^{\rm SI}_{n}}}
\newc{\sigsdp}{\ensuremath{\sigma^{\rm SD}_{p}}}	\newc{\sigsdn}{\ensuremath{\sigma^{\rm SD}_{n}}}
\newc{\sigsi}{\ensuremath{\sigma^{\rm SI}}}	\newc{\sigsd}{\ensuremath{\sigma^{\rm SD}}}

\newc{\abund}{\ensuremath{ \Omega h^2}}
\newc{\omegadm}{\ensuremath{ \Omega_{{\rm DM}}}}     \newc{\abunddm}{\ensuremath{ \Omega_{{\rm DM}} h^2}} 
\newc{\omegam}{\ensuremath{ \Omega_{{\rm m}}}}       \newc{\abundm}{\ensuremath{ \Omega_{{\rm m}} h^2}}
\newc{\omegab}{\ensuremath{ \Omega_{{\rm b}}}}	\newc{\abundb}{\ensuremath{ \Omega_{{\rm b}} h^2}}
\newc{\omegatot}{\ensuremath{ \Omega_{{\rm TOT}}}}
\newc{\omegacdm}{\ensuremath{ \Omega_{{\rm CDM}}}}   \newc{\abundcdm}{\ensuremath{ \Omega_{{\rm CDM}} h^2}}
\newc{\omegalambda}{\ensuremath{ \Omega_{\Lambda}}} \newc{\abundlambda}{\ensuremath{ \Omega_{\Lambda} h^2}}
\newc{\omegarad}{\ensuremath{ \Omega_{{\rm rad}}}}  \newc{\abundrad}{\ensuremath{ \Omega_{{\rm rad}} h^2}}

\newc{\rhocrit}{\ensuremath{ \rho_{\rm crit}}}
\newc{\rhochi}{\ensuremath{ \rho_{\chi}}}

\newc{\abunchi}{\ensuremath{\Omega_\chi h^2}}
\newc{\abundlsp}{\ensuremath{\Omega_{\rm LSP}h^2}}

\newc{\amu}{\ensuremath{ a_{\mu}}}        \newc{\amususy}{\ensuremath{ a_{\mu}^{\mathrm{SUSY}}}}
\newc{\amuexpt}{\ensuremath{ a_{\mu}^{\mathrm{expt}}}}        \newc{\amusm}{\ensuremath{ a_{\mu}^{\mathrm{SM}}}}
\newc\deltaamu{\ensuremath{\Delta a_{\mu}}} \newc{\deltaamususy}{\ensuremath{\delta a_{\mu}^{\mathrm{SUSY}}}}
\newc\gmtwo{\ensuremath{ (g-2)_{\mu}}} 
\newc{\deltagmtwomususy}{\ensuremath{\delta\left(g-2\right)_{\mu}^{\mathrm{SUSY}}}}
\newc{\deltagmtwomu}{\ensuremath{\delta\left(g-2\right)_{\mu}}}

\newc\BR{\ensuremath{\rm BR}}

\newc\bsgamma{\ensuremath{ b\rightarrow s \gamma }}
\newc\bxsgamma{\ensuremath{\overline{B}\rightarrow X_{s}\gamma}}

\newc\brbsgamma{\ensuremath{\BR\left(\bsgamma\right)}}
\newc\brbxsgamma{\ensuremath{\BR\left(\bxsgamma\right)}}

\newc\bsmumu{\ensuremath{B_s\to\mu^+\mu^-}}
\newc\brbsmumu{\ensuremath{\BR\left(B_s\to\mu^+\mu^-\right)}}

\newc\bdmmumu{\ensuremath{\overline{B}_d\to\mu^+\mu^-}}

\newc\bbbarmix{\ensuremath{\overline{B}_s\mbox{-}B_s}}      
\newc\delmbs{\ensuremath{\Delta M_{B_s}}}

\newc{\butaunu}{\ensuremath{B_u \rightarrow \tau \nu}}
\newc{\brbutaunu}{\ensuremath{\BR\left(B_u \rightarrow \tau \nu\right)}}

     \newcommand*{\refsec}[1]{Sec.~\ref{#1}}

\newcommand*{\neutone}{\ensuremath{\chi}}

\newcommand*{\micromegas}{MicrOMEGAs}

\newcommand*{\superiso}{\text{SuperIso}}

\newcommand*{\nmssmtools}{\text{NMSSMTools}}

\let\oldcite\cite
\renewcommand*{\cite}{~\oldcite}

\newcommand*{\hl}{\ensuremath{h}}
